\documentclass{article}

\usepackage{arxiv}

\usepackage[utf8]{inputenc} 
\usepackage[T1]{fontenc}    
\usepackage{hyperref}       
\usepackage{url}            
\usepackage{booktabs}       
\usepackage{amsfonts}       
\usepackage{nicefrac}       
\usepackage{microtype}      
\usepackage{lipsum}
\usepackage{graphicx}
\usepackage{amsmath}
\usepackage{comment}
\usepackage{caption}
\usepackage{multirow} 
\graphicspath{ {./images/} }

\title{A Comprehensive Survey of Linear, Integer, and Mixed-Integer Programming Approaches for Optimizing Resource Allocation in 5G and Beyond Networks}

\author{
 Naveed Ejaz \\
  School of Computing\\
  Queen's University, \\
  Kingston, ON, Canada \\
  \texttt{naveed.ejaz@queensu.ca} \\
   \And
 Salimur Choudhury \\
School of Computing\\
  Queen's University, \\
  Kingston, ON, Canada \\
  \texttt{s.choudhury@queensu.ca} \\
  }

\begin{document}
\maketitle
\begin{abstract}
The introduction of 5G networks has significantly advanced communication technology, offering faster speeds, lower latency, and greater capacity. This progress sets the stage for Beyond 5G (B5G) networks, which present new complexity and performance requirements challenges. Linear Programming (LP), Integer Linear Programming (ILP), and Mixed-Integer Linear Programming (MILP) models have been widely used to model the optimization of resource allocation problems in networks. This paper reviews 103 studies on resource allocation strategies in 5G and B5G, focusing specifically on optimization problems modelled as LP, ILP, and MILP. The selected studies are categorized based on network architectures, types of resource allocation problems, and specific objective functions and constraints. The review also discusses solution methods for NP-hard ILP and MILP problems by categorizing the solution methods into different categories. Additionally, emerging trends, such as integrating AI and machine learning with optimization models, are explored, suggesting promising future research directions in network optimization. The paper concludes that LP, ILP, and MILP models have been widely adopted across various network architectures, resource types, objective functions, and constraints and remain critical to optimizing next-generation networks. 
\end{abstract}

\keywords{5G Networks \and Beyond 5G Networks \and Resource Allocation \and Linear Programming (LP) \and Integer Linear Programming (ILP) \and Mixed-Integer Linear Programming (MILP) \and Network Optimization \and Quality of Service (QoS)}

\section{Introduction}
The advent of 5G networks has brought in a new era of communication technology characterized by unprecedented speed, lower latency, and higher capacity \cite{9355403}. This technological evolution promises to support various applications, from Enhanced Mobile Broadband (eMBB) to mission-critical communications and massive Internet of Things(IoT). As we transition from 5G to the more advanced Beyond 5G (B5G) networks, the systems become increasingly complex and demanding \cite{mahmood2021industrial, mishra2020survey}. 

Resource allocation in 5G and B5G networks involves the dynamic and efficient distribution of network resources (such as spectrum, power, and network infrastructure) to meet diverse and evolving service requirements while ensuring optimal network performance, user satisfaction, and energy efficiency \cite{ebrahimi2024resource}. The growing complexity of these networks and the diverse applications and services they support require sophisticated optimization strategies to effectively manage their performance and efficiency.

Researchers have extensively explored the application of mathematical optimization techniques, particularly Linear Programming (LP), Integer Linear Programming (ILP), and Mixed-Integer Linear Programming (MILP), for resource allocation in 5G and B5G networks \cite{fayad2024optimization}.   These techniques offer a powerful and flexible framework for modelling and solving complex optimization problems, enabling the efficient allocation of resources while considering various constraints and objectives \cite{machiwa2024comprehensive}. The existing literature on resource allocation in 5G and B5G networks is vast and diverse, encompassing many network architectures, resource types, and optimization goals \cite{xu2021survey}. However, a comprehensive survey that consolidates and synthesizes the research specifically within the LP, ILP, and MILP frameworks is lacking. This paper addresses this gap by reviewing 103 studies on resource allocation strategies in 5G and B5G networks that utilize LP, ILP, and MILP models.

To establish the widespread usage of these models, we categorize the 103 selected studies based on network architecture types, resource allocation types, common objective functions, and the types of constraints considered. We conclude that LP, ILP, and MILP have been extensively employed in modelling optimization problems in 5G and B5G networks, and they remain relevant and continue to play a crucial role in modelling resource allocation challenges in next-generation networks.

The survey also provides a detailed analysis of the solution techniques and approaches for solving LP, ILP, and MILP problems in network optimization. The paper discusses solution techniques such as direct LP formulations, transformations to LP, and the direct use of solvers for solving ILP and MILP problems. Given the NP-hard nature of ILP and MILP \cite{ganian2017going} and the limitations of ILP and MILP solvers in addressing larger resource allocation problems, the literature proposes approximation and heuristic algorithms. We classify and categorize these solutions into multiple categories, including exact algorithms with heuristic components, metaheuristics, local search and constructive heuristics, hybrid heuristics and reinforcement learning, approximation algorithms, decomposition-based heuristics, heuristic-based relaxations, and problem-specific heuristics.

The survey concludes by highlighting the promising future research directions in network optimization. It emphasizes the integration of AI and machine learning with optimization models, suggesting that this synergy will play a pivotal role in shaping the future of resource allocation strategies in next-generation networks.

\begin{figure*}[t]
    \centering
    \includegraphics[scale=0.42]{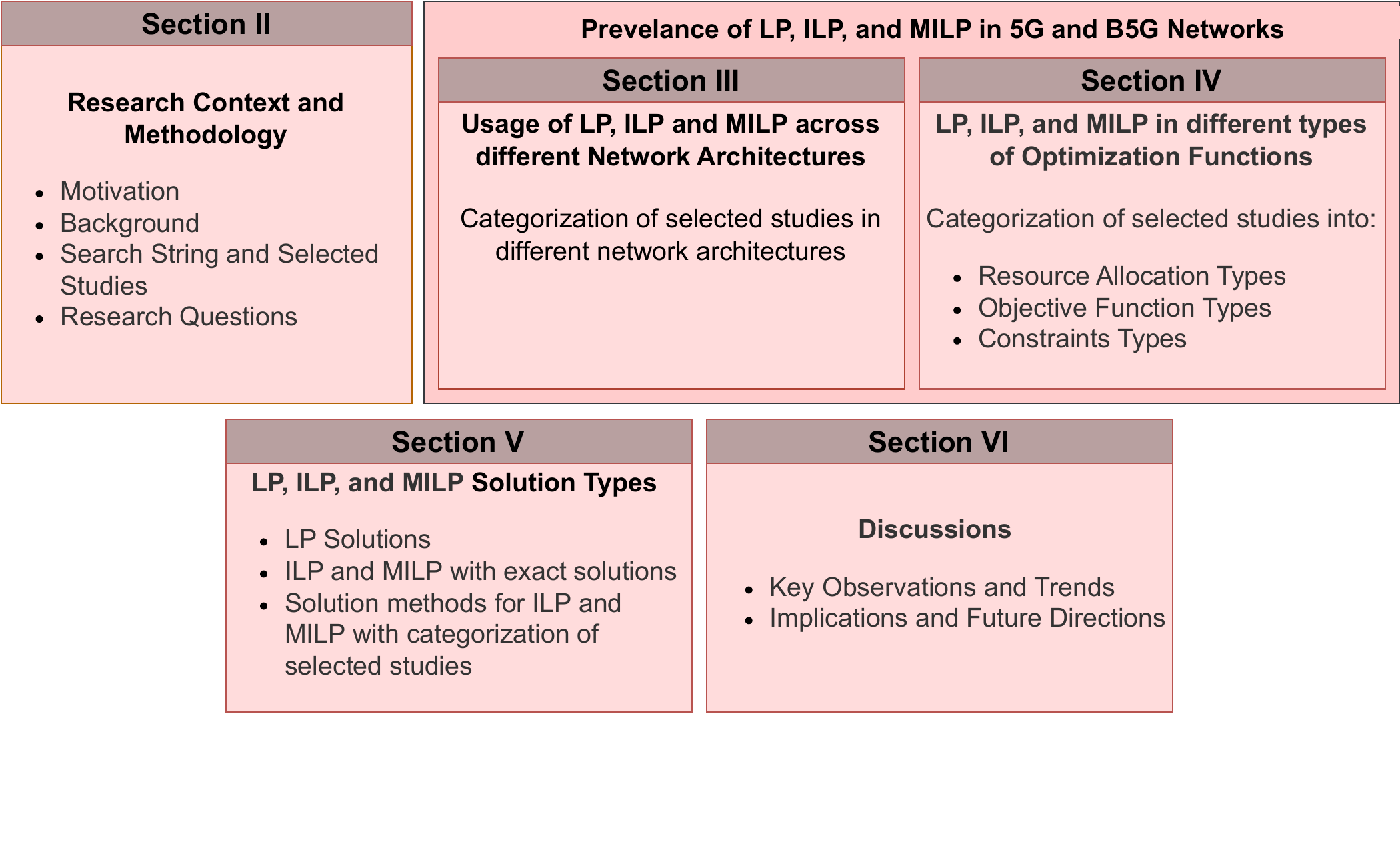}
    \caption{Structure of the paper}
    \label{fig:structure}
\end{figure*}

Figure \ref{fig:structure} shows the arrangement of significant sections of the paper. Section \ref{sec:BG} establishes the survey context by highlighting the research's motivation, providing background, and outlining the methodological approach for selecting and analyzing relevant studies. Following this, Section \ref{sec:NW_Arch_Objective_Funct} examines the prevalence of LP, ILP, and MILP models across various network architectures, demonstrating their adaptability and effectiveness in addressing resource allocation challenges in different network types. Section \ref{sec:Optimization} explores the application of these models to various optimization functions, resource types, and constraint types, emphasizing their role in optimizing network performance and resource utilization efficiency. Section \ref{Solution_Techniques} provides a detailed analysis of the solution techniques and approaches for solving LP, ILP, and MILP problems in network optimization. Section \ref{sec:Discussion} synthesizes the key findings of the survey and proposes potential future research directions in network optimization. Finally, Section \ref{sec:Conclusions} wraps up the paper with concluding remarks.

\section{Research Context and Methodology}
\label{sec:BG}

This section lays the foundation for our comprehensive survey by presenting the motivation behind the research, providing background information, and detailing the methodological approach adopted.

\subsection{Motivation}
\label{subsec:Motivation}

\noindent
The transition from 4G to 5G and beyond has significantly complicated network architectures with features like network slicing \cite{arnhold2024network}, massive MIMO(Multiple Input Multiple Output) \cite{elijah2022intelligent}, and Ultra-Reliable Low Latency Communication (URLLC) \cite{li20185g}. These complexities require sophisticated resource allocation strategies for effectively managing spectrum, power, and computational resources. Mathematical optimization techniques such as LP, ILP, and MILP offer rigorous and flexible solutions for these challenges. 

Despite numerous studies on network optimization, a comprehensive survey that consolidates research within the LP, ILP, and MILP frameworks, specifically for 5G and future network technologies, is needed. In this context, this survey serves as a valuable reference, helping researchers fully understand the range of mathematical tools and their practical implications. It details how LP, ILP, and MILP techniques are solved to meet specific service requirements, highlighting scalability as networks grow.

Furthermore, ILP and MILP face challenges in computational efficiency in large-scale deployments \cite{mirhosseini2023fast}. This survey documents these issues and discusses the recent advancements in computational techniques like decomposition methods, parallel processing, and heuristics. Lastly, the survey identifies emerging areas and trends, providing direction for future research and development in network resource allocation.


\subsection{Background}
\label{subsec:Background}

\noindent
LP is a mathematical optimization technique used to find the best outcome in a mathematical model whose requirements are represented by linear relationships. The objective function, representing the quantity to be optimized, and the constraints, representing limitations or requirements, are all linear functions of the decision variables. The objective function in LP is to maximize (or minimize): $Z = c_1x_1 + c_2x_2 + \ldots + c_nx_n$, where $c_i$ are the coefficients of the objective function, and $x_i$ are the decision variables. The constraints can be represented as:

\begin{equation}
\begin{split}
a_{11}x_1 + a_{12}x_2 + \ldots + a_{1n}x_n &\leq (\geq,=) b_1 \\
a_{21}x_1 + a_{22}x_2 + \ldots + a_{2n}x_n &\leq (\geq,=) b_2 \\
&\vdots \\
a_{m1}x_1 + a_{m2}x_2 + \ldots + a_{mn}x_n &\leq (\geq,=) b_m
\end{split}
\end{equation}


\noindent
where \(x_1, x_2, \ldots, x_n \in \mathbb{Z}\), $a_{ij}$ represents the coefficients of the constraints, $b_i$ are the right-hand side values of the constraints, $m$ is the number of constraints and $n$ is the number of decision variables.

\noindent
ILP is a mathematical optimization problem where all variables are integers, but the objective function and constraints are linear. The objective function $Z$ aims to maximize or minimize: $Z = c_1x_1 + c_2x_2 + \ldots + c_nx_n$. In this objective function, $c_i$ are the coefficients of the objective function and $x_i$ are the decision variables. The constraints are given as: 

\begin{equation}
\begin{aligned}
a_{11}x_1 + a_{12}x_2 + \ldots + a_{1n}x_n &\leq (\geq,=) b_1 \\
a_{21}x_1 + a_{22}x_2 + \ldots + a_{2n}x_n &\leq (\geq,=) b_2 \\
&\vdots \\
a_{m1}x_1 + a_{m2}x_2 + \ldots + a_{mn}x_n &\leq (\geq,=) b_m
\end{aligned}
\end{equation}



\noindent
MILP combines ILP and LP by allowing integer and continuous variables, making it versatile for real-world optimization problems. The objective function \(Z\) in a MILP aims to maximize or minimize: $ Z = c_1 x_1 + c_2 x_2 + \ldots + c_n x_n + d_1 y_1 + d_2 y_2 + \ldots + d_m y_m $, where \(c_i\) and \(d_j\) are the coefficients for the decision variables \(x_i\) and \(y_j\) respectively. Here, \(x_i\) are the integer variables, and \(y_j\) are the continuous variables. The constraints in a MILP model are given as:

\begin{equation}
\begin{split}
\begin{aligned}
&a_{11}x_1 + a_{12}x_2 + \ldots + a_{1n}x_n + a'_{11}y_1 + a'_{12}y_2 + \ldots + a'_{1m}y_m \\
&\quad\leq (\geq,=) b_1 \\
&a_{21}x_1 + a_{22}x_2 + \ldots + a_{2n}x_n + a'_{21}y_1 + a'_{22}y_2 + \ldots + a'_{2m}y_m \\
&\quad\leq (\geq,=) b_2 \\
&\vdots \\
&a_{k1}x_1 + a_{k2}x_2 + \ldots + a_{kn}x_n + a'_{k1}y_1 + a'_{k2}y_2 + \ldots + a'_{km}y_m \\
&\quad\leq (\geq,=) b_k
\end{aligned}
\end{split}
\end{equation}

\noindent
where \(x_1, x_2, \ldots, x_n \in \mathbb{Z}\) (integer variables) and \(y_1, y_2, \ldots, y_m\) are continuous variables (either \(y_j \geq 0\) or unrestricted).

LP, ILP, and MILP are widely used to model network optimization problems. LP is ideal for optimization problems in which the resources can be divided into fractions \cite{smith2012measuring}. LP optimization problems can be efficiently solved in large-scale networks using simplex or interior point techniques \cite{bixby1992very}. However, many network resources, such as routers, switches, channels, base stations, and servers, can not be divided. ILP is designed for situations where decisions are binary (e.g., whether to include a link) or resources are counted in whole numbers (e.g., the number of servers or base stations). ILP models complex constraints and objectives, such as capacity limits, Quality of Service (QoS) requirements, and topological constraints. MILP combines the strengths of LP and ILP by allowing both continuous and discrete variables in the same model \cite{wiese2016interplay}. This is ideal for network resource allocation, where resources like bandwidth can be allocated continuously, while others like routers or wavelengths must be allocated in discrete units. MILP can model many optimization problems, from network design and capacity planning \cite{baghestani2023logistics} to traffic engineering \cite{amaldi2011milp} and energy-efficient routing \cite{meng2020milp}. Its versatility enables the formulation of complex optimization problems that closely match real-world scenarios. 

\subsection{Research Questions}
Following research questions are addressed in this paper: 
\begin{itemize}
    \item RQ 1: How prevalent are LP, ILP, and MILP in modelling resource allocation problems across different network architectures?
    \item RQ 2: What is the prevalence of LP, ILP, and MILP in resource allocation problems and different types of objective functions and constraints?
    \item RQ 3: What are the most commonly used solution techniques and approaches for solving LP, ILP, and MILP problems in network optimization?
    \item RQ4: What are the future directions in designing and solving optimization problems using LP, ILP, and MILP? 
\end{itemize}

\subsection{Search String and Databases Searched}
\label{subsec:SearchString}
\noindent
We queried several research databases, including ACM Digital Library, IEEE Xplore, Scopus, and Web of Science, using the following search query: (5G OR Beyond 5G OR B5G OR next-generation networks) AND (resource allocation OR network optimization) AND (linear programming OR LP OR integer linear programming OR ILP OR mixed-integer linear programming OR MILP). After the classification of resource allocation types was identified, the second phase of the search was carried out using the following nine sub-strings: (5G OR Beyond 5G OR B5G OR next-generation networks) AND (Resource Type) AND (linear programming OR LP OR integer linear programming OR ILP OR mixed-integer linear programming OR MILP), where resource type is replaced by one of the nine identified resource allocation types:  Spectrum Allocation, Radio Resource Management, Computing Resources Allocation, Network Slicing, Energy Resource Management, Backhaul and Fronthaul Resources Allocation, Storage Resources Allocation, QoS and Quality of Experience (QoE) Management, User and Device Connectivity Management. In the first screening phase, we applied the selection criteria based on titles and abstracts to filter out irrelevant studies. In the second phase, we applied the criteria again using the full text. After applying all the screening stages, 103 papers were selected.

\section{Prevalence of LP Variants in Diverse Network Architectures}
\label{sec:NW_Arch_Objective_Funct}

\captionsetup{justification=centering}

\begin{figure*}[htbp]
    \centering
     \includegraphics[width=0.8\linewidth]{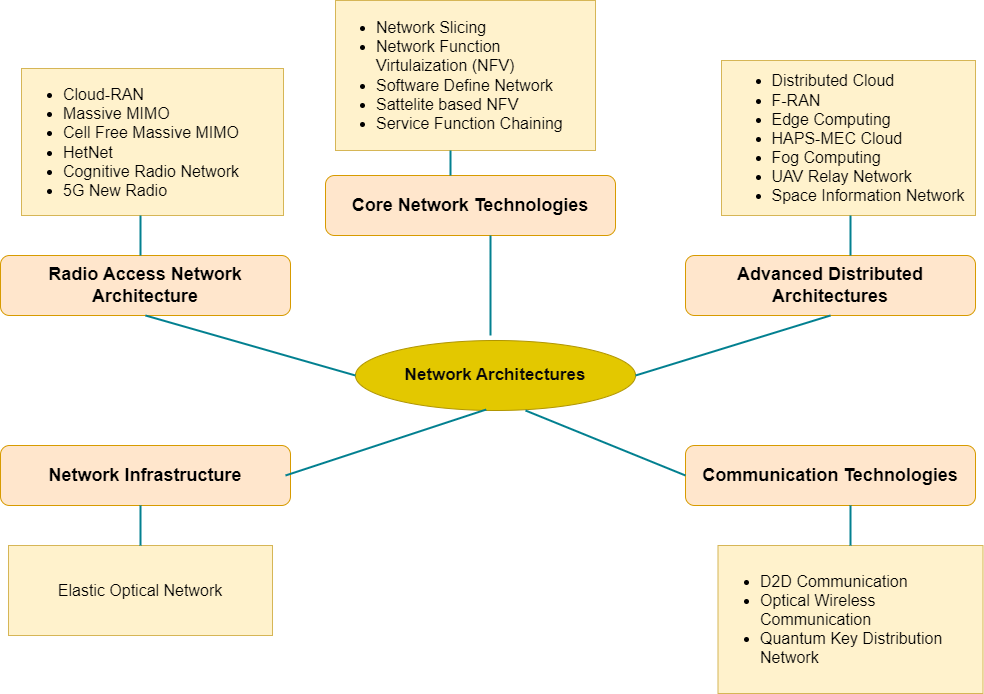}
    \caption{Categorization of Network Architectures}
    \label{fig:nw-arch}
\end{figure*}



\noindent

Resource allocation problems have been modelled using LP, ILP, and MILP across various network architectures in 5G and B5G literature. This section highlights the widespread application of LP variants in resource allocation across different network architectures. The selected studies are categorized based on the relevant network architectures they address. Figure \ref{fig:nw-arch} shows the classification of network architecture types in studies that utilized LP, ILP, and MILP for resource allocation. 


Table \ref{Tab-NW-Arch-Categ} categorizes selected studies based on the network architecture they focused on and lists the corresponding references that utilized LP, ILP, and MILP for resource allocation optimization. The table demonstrates that resource allocation modelling using LP, ILP, or MILP is a widely adopted approach across various network architectures. Next, we briefly discuss the main network architectures in which resource allocation problems have been modelled as LP, ILP, and MILP. 


Radio Access Network (RAN) architectures are essential for connecting devices to networks via radio connections, evolving from 4G to 5G to meet demands for higher data rates, lower latency, and better efficiency \cite{marsch20165g,habibi2019comprehensive}. Modern RAN technologies include Cloud-RAN(C-RAN), which centralizes base station processing in cloud data centers, improving resource allocation, power efficiency, and system capacity through advanced techniques like Coordinated Multi-Point (CoMP) \cite{rodoshi2020resource}. Several of the studies selected for this review on C-RAN focus on optimizing resource allocation and reducing energy consumption by incorporating cloud-based models and optical fronthaul solutions \cite{hasabelnaby2018joint, sharara2023minimizing, gkatzios2018compute, gkatzios2019efficiency, de2019optimal, ali2017joint, fayad2022design}. Additionally, emerging concepts such as Open RAN (O-RAN) and network slicing, highlighted in the reviewed studies, aim to minimize delays and enhance resource management \cite{kazemifard2021minimum, 9827120}. Lastly, the selected studies on virtualization in C-RAN emphasize dynamic resource allocation and Virtual Network Functions (VNFs) placement \cite{tang2018queue, mharsi2018joint, arouk2017multi, klinkowski2022dedicated, basu2020reverse}.

Massive MIMO systems use hundreds of antennas at base stations to boost cell capacity, spectral efficiency, and signal quality through beamforming but face challenges related to signal processing complexity and antenna array size \cite{mounika2021downlink}. The selected studies on Massive MIMO include \cite{xue2023joint}, which proposed a resource allocation strategy in a cell-free Massive MIMO system enhanced by Time and Wavelength Division Multiplexing Passive Optical Network (TWDM-PON); \cite{feng2017adaptive}, which focused on pilot optimization and resource allocation in Heterogeneous Networks (HetNet); \cite{feng2017joint}, which explored joint frame design for maximizing downlink rates; and \cite{mounika2021downlink}, which discussed downlink resource allocation.

 HetNets use various cell types (macro, micro, pico, femtocells) to optimize coverage and efficiency while managing interference with Inter-Cell Interference Coordination (ICIC) techniques \cite{zoha2017leveraging, you2017load}. The selected studies include research on patient-centric 6G HetNets for e-healthcare \cite{hadi2020patient}, Massive MIMO for frame design and pilot optimization \cite{feng2017adaptive, feng2017joint}, and distributed HetNet architectures for dynamic channel allocation \cite{swetha2018dynamic}. Additionally, energy efficiency improvements and resource allocation algorithms for 5G LTE-based HetNets \cite{xie2017energy, saghezchi2017energy, liu2020joint, jain2021user}, as well as CRAN principles combined with HetNet structures \cite{ali2017joint}, have been explored. Data-driven approaches include interference-aware strategies using Call Detail Records (CDR) data \cite{zoha2017leveraging} and load optimization in LTE networks \cite{you2017load}. 

5G-New Radio (5G-NR) is a flexible air interface for 5G networks, supporting wider bandwidths, higher frequencies, and various connection types \cite{jalali2020joint}. The selected studies for review encompass a range of innovative approaches. They include a scalable physical layer framework with mini-slots \cite{marijanovic2020multiplexing}, optimized user association and power allocation strategies for millimetre-wave communication \cite{fayad20235g}, and advancements in error protection through polar codes and adaptive codebooks for Sparse Code Multiple Access (SCMA) \cite{jalali2020joint, zhai2017adaptive}. Cognitive radio research highlights opportunistic scheduling and spectral-energy efficiency trade-offs \cite{salameh2022opportunistic, sasikumar2021spectral}. Energy efficiency improvements are explored through new RAN architectures and relay selection methods \cite{zorello2022power, saghezchi2017energy, sharara2022policy}. Additionally, deployment strategies involving UAVs as relays and Device-to-Device (D2D) communication integration are examined \cite{javad2021re, alwan2018joint, yoon2018resource, vlachos2017moca}, alongside dynamic channel allocation in small cells and packet-optical transport networks \cite{swetha2018dynamic, klinkowski2022dedicated}.

In a distributed cloud architecture, computing resources are spread across multiple locations to reduce latency and backhaul traffic, crucial for 5G applications like autonomous driving and real-time analytics \cite{8676260}.  From the selected studies, we discuss research on using a shared O-Cloud (Open RAN Cloud) with machine learning for dynamic adaptation \cite{sharara2022policy}, integrating RAN and core network components for dynamic network slicing \cite{8676260}, and implementing network flying platforms as aerial hubs \cite{alsheyab2019near}. A model incorporating Electromagnetic Fields (EMF) constraints in 5G network planning within a distributed cloud framework has also been proposed \cite{chiaraviglio20215g}.

Fog radio access network (F-RAN) integrates fog computing principles into the RAN, decentralizing computing resources to the network edge to reduce latency, enable local content caching, and improve responsiveness \cite{zeng2019energy}. From the selected studies, we discuss research on an F-RAN architecture that combines fog and cloud computing to enhance energy efficiency \cite{zeng2019energy}, an NFV-based architecture aimed at reducing energy consumption \cite{al2019optimized}, and applications in gaming for real-time processing \cite{spinelli2022migration}. Additionally, hierarchical network slicing and power-efficient beam allocation in IoT networks have been explored \cite{chu2023hirn, deb2022fog}. 

EdgeNet, or edge computing, processes data closer to its source, reducing latency and congestion, and supporting real-time analytics, AI, and IoT \cite{gatzianas2021energy, spinelli2022migration, lagkas2021optimized}. From the selected studies, we discuss research on VNF placement strategies aimed at reducing power consumption \cite{gatzianas2021energy}, resource crowdsourcing frameworks \cite{pan2020multi}, and Mobile Edge Computing (MEC) deployment in 5G networks \cite{lagkas2021optimized}. These studies also cover enhancements in gaming quality \cite{spinelli2022migration}, optimization of caching \cite{ge2021dynamic}, and improvements in edge computing migrations \cite{kim2022modems}. Emerging technologies explored include UAV network reconfiguration \cite{javad2021re} and fog-based power-efficient beam allocation \cite{deb2022fog}. While these advancements offer reduced latency and improved scalability, they also present challenges in resource management, security, and scalability \cite{jia2021toward, deb2022fog}.

Among various network architectures, integrating High Altitude Platform Stations (HAPS) with MEC and cloud infrastructure creates a layered system that provides high-speed internet over extensive areas, including remote regions \cite{jia2021toward, yang2020towards}. Unmanned Aerial Vehicle (UAV) Relay Networks utilize drones as aerial relay stations to enhance communication coverage, particularly in remote or compromised areas. These drones extend network access and improve coverage in 5G and B5G networks by enabling dynamic, multi-hop communication \cite{javad2021re}. The Space Information Network (SIN) integrates satellite technology into communication networks for global coverage, using satellites to extend access to remote areas, which is crucial for 5G and B5G networks. Research in this area includes optimizing VNFs deployment and flow routing within SIN \cite{yang2023space} and enhancing data collection with Space-Air-Ground Integrated Networks (SAGIN) involving HAPs and Low Earth Orbit (LEO) satellites \cite{jia2021toward}. These studies have been selected for inclusion in the survey paper due to their relevance and contribution to the field.

The prevalence of LP, ILP, and MILP across a broad spectrum of network architectures makes these techniques worthy of study. It confirms their ongoing relevance in addressing the evolving challenges of modern network systems. The examples from Table \ref{Tab-NW-Arch-Categ} show that both established and emerging network architectures continue to rely on these optimization methods, making them indispensable tools for researchers and practitioners working on resource allocation in various network contexts.



\begin{table*}[ht]
\centering
\caption{Categorization of Selected Studies based on Network Architecture}
\label{Tab-NW-Arch-Categ}
\begin{tabular}{p{0.2\textwidth} p{0.6\textwidth}}
\hline
\textbf{Network Architecture} & \textbf{References} \\
\hline
C-RAN & \cite{hasabelnaby2018joint}, \cite{sharara2023minimizing}, \cite{gkatzios2018compute}, \cite{gkatzios2019efficiency},  \cite{de2019optimal},  \cite{ali2017joint}, \cite{fayad2022design}, \cite{kazemifard2021minimum}, \cite{9827120}, \cite{tang2018queue},   \cite{mharsi2018joint}, \cite{arouk2017multi}, \cite{klinkowski2022dedicated}, \cite{basu2020reverse} \\
\hline
Massive MIMO & \cite{mounika2021downlink}, \cite{xue2023joint}, \cite{feng2017adaptive}, \cite{feng2017joint} \\
\hline
HetNet & \cite{ali2017joint}, \cite{feng2017adaptive}, \cite{feng2017joint}, \cite{zoha2017leveraging}, \cite{you2017load},  \cite{hadi2020patient}, \cite{swetha2018dynamic}, \cite{monteiro2018distributed}, \cite{xie2017energy}, \cite{saghezchi2017energy},  \cite{liu2020joint}, \cite{jain2021user}, \cite{saeed2016control} \\
\hline
5G-NR architecture & \cite{jalali2020joint}, \cite{zhai2017adaptive}, \cite{salameh2022opportunistic}, \cite{sasikumar2021spectral}, \cite{zorello2022power},  \cite{sharara2022policy}, \cite{javad2021re}, \cite{alwan2018joint}, \cite{yoon2018resource},  \cite{guo2023delay}, \cite{swetha2018dynamic}, \cite{saghezchi2017energy}, \cite{zoha2017leveraging}, \cite{vlachos2017moca}, \cite{marijanovic2020multiplexing}, \cite{fayad20235g}, \cite{deb2022fog} \\
\hline
Distributed Cloud & \cite{8676260}, \cite{alsheyab2019near}, \cite{chiaraviglio20215g}, \cite{sharara2022policy} \\
\hline
F-RAN & \cite{deb2022fog}, \cite{zeng2019energy}, \cite{al2019optimized}, \cite{chu2023hirn}, \cite{spinelli2022migration} \\
\hline
Edge Computing & \cite{gatzianas2021energy}, \cite{spinelli2022migration}, \cite{lagkas2021optimized}, \cite{spinelli2022edge}, \cite{ge2021dynamic}, \cite{kim2022modems}, \cite{guo2023delay}, \cite{yang2023space}, \cite{gao2019deep}, \cite{yue2023delay}, \cite{javad2021re}, \cite{jia2021toward}, \cite{deb2022fog}, \cite{prastowo2022migration}, \cite{pan2020multi} \\
\hline
HAPS-MEC-Cloud system & \cite{yang2020towards}, \cite{jia2021toward} \\
\hline
UAV Relay Network & \cite{javad2021re} \\
\hline
SIN & \cite{yang2023space}, \cite{jia2021toward} \\
\hline
Network Slicing & \cite{ko2022pdras}, \cite{ko2021priority}, \cite{li2019latency}, \cite{boutiba2023optimal}, \cite{mei20205g}, \cite{harutyunyan2019orchestrating}, \cite{yang2019two}, \cite{fendt2018formal}, \cite{fendt2018network}, \cite{jia2017efficient}, \cite{papagianni2018rethinking}, \cite{de2020optimal}, \cite{al2019optimized}, \cite{morais2022placeran}, \cite{bi2019resource}, \cite{yu2019network}, \cite{biallach2022optimization}, \cite{mushtaq2023optimal}, \cite{gholami2023mobile}, \cite{9379507}, \cite{luu2021uncertainty}, \cite{10056963}, \cite{9929619}, \cite{9827120}, \cite{9685847}, \cite{8676260}, \cite{ferdosian20225g}, \cite{you2018resource}, \cite{spinelli2022migration}, \cite{pan2020multi}, \cite{prastowo2022migration}, \cite{ge2021dynamic}, \cite{gatzianas2021offline}, \cite{erbati2021application}, \cite{gang2019inter} \\
\hline
Network Function Virtualization & \cite{pham2021traffic}, \cite{spinnewyn2018coordinated}, \cite{you2017load}, \cite{quang2018qaav}, \cite{erbati2021application}, \cite{javad2021re}, \cite{yang2023space}, \cite{almasaeid2023minimum}, \cite{yue2023delay} \\
\hline
Software Defined Network(SDN) & \cite{8204056}, \cite{xie2017energy}, \cite{zoha2017leveraging}, \cite{quang2018qaav}, \cite{almasaeid2023minimum}, \cite{erbati2021application}, \cite{javad2021re}, \cite{yang2023space}, \cite{jia2021toward} \\
\hline
D2D Communication &  \cite{alwan2018joint}, \cite{vlachos2017moca}, \cite{yoon2018resource}, \cite{al2021joint} \\
\hline
Optical Wireless Communication System & \cite{aletri2020effect}, \cite{aletri2020optimum}, \cite{cao2021hybrid} \\
\hline
EON & \cite{cao2021hybrid}, \cite{kokkinos2019pattern} \\
\hline
\end{tabular}
\end{table*}

\section{Prevelance of LP, ILP, and MILP Across Various Types of Optimization Functions}
\label{sec:Optimization}

This section provides a comprehensive overview of the prevalence and application of LP, ILP, and MILP across various optimization functions within 5G and B5G networks. We categorize the selected studies into three key dimensions: resource types, constraint types, and objective function types. By examining these dimensions, we aim to highlight the specific roles that LP, ILP, and MILP play in optimizing network operations, ensuring efficient resource allocation, and meeting the stringent performance requirements of next-generation networks.

\subsection {Resource allocation types}

The resources discussed in the selected studies can be broadly classified into several types. Table \ref{Tab:Resouce_Types} illustrates that several studies concurrently addressed multiple resource allocation problems. 


Among the selected studies, some studies explored spectrum allocation in 5G and B5G networks, addressing frequency band distribution \cite{boutiba2023optimal, aletri2020effect}, Cognitive Radio (CR) capabilities \cite{salameh2022opportunistic}, and dynamic channel allocation  \cite{swetha2018dynamic, almasaeid2023minimum, saeed2016control}. These models aimed to optimize performance and reduce interference in various contexts, including LTE, IoT networks, and UAV-assisted systems.

Resource allocation for network slicing focuses on managing multiple logical networks over a shared infrastructure. Studies discuss dynamic resource management \cite{ko2022pdras}, optimal VNF deployment \cite{de2019optimal}, and the integration of slicing information \cite{gatzianas2021offline} to enhance performance in 5G and beyond.

    Radio Resource Management (RRM) is essential for optimizing wireless communication by managing radio frequency resources to ensure fairness and QoS. The research addressed RRM techniques in contexts like Non-Orthogonal Multiple Access (NOMA)  \cite{marijanovic2020multiplexing}, Massive MIMO \cite{mounika2021downlink}, and hybrid cloud infrastructures \cite{de2020optimal}, emphasizing power control and spectrum efficiency \cite{zorello2022power}.

    In Network Function Virtualization (NFV) and Software Defined Network(SDN) frameworks, computing resource allocation enhances network flexibility and efficiency. Studies discuss allocation strategies for Virtual Machines (VMs) \cite{prastowo2022migration}, Containers \cite{gatzianas2021energy}, MEC nodes \cite{lagkas2021optimized, spinelli2022edge}, and computational resources \cite{sharara2023minimizing, de2019optimal}, focusing on real-time management across physical and virtual devices.

Energy Resource Management (ERM) involves the strategic management of both renewable and non-renewable energy sources to ensure efficiency and sustainability. The focus is balancing variable renewable energies with stable sources to maintain grid stability and optimize energy use \cite{spinelli2022migration, spinelli2022edge}.

     Research on QoS and QoE emphasizes efficient resource allocation to meet specific service quality requirements of different applications, ensuring high user satisfaction in 5G networks through optimized dual connectivity  \cite{fayad20235g}, power allocation, and relay node selection \cite{saghezchi2017energy}.

   Effective resource allocation for backhaul and fronthaul in 5G and B5G networks is critical for maintaining network performance and scalability. Studies focus on managing interference \cite{zoha2017leveraging}, optimizing Base Band Units (BBUs) functions \cite{gkatzios2019efficiency}, and minimizing energy consumption \cite{zoha2017leveraging} in these network segments.

    In MEC architecture, resource allocation at the network edge reduces latency and improves service quality. Studies discuss the allocation of computation \cite{spinelli2022migration, pan2020multi, prastowo2022migration, gatzianas2021energy, harutyunyan2019orchestrating, bi2019resource}, energy \cite{spinelli2022migration, pan2020multi, prastowo2022migration, gatzianas2021energy, bi2019resource}, network, communication  \cite{spinelli2022migration, prastowo2022migration, gatzianas2021energy, harutyunyan2019orchestrating, bi2019resource}, and storage resources \cite{pan2020multi, harutyunyan2019orchestrating}to enhance edge computing capabilities.

URLLC requires high reliability and low latency for critical applications. The research addressed the allocation of backhaul \cite{alsheyab2019near} and fronthaul \cite{lagkas2021optimized} resources, focusing on optimizing BBU functions \cite{gkatzios2019efficiency} and managing interference to meet the stringent requirements of URLLC services \cite{zoha2017leveraging}.

Resource allocation for mobility management is crucial to maintaining service quality in wireless networks. Studies explored strategies for optimizing service migrations \cite{prastowo2022migration}, bandwidth allocation, and route reconfiguration in 4G LTE and 5G networks \cite{javad2021re, papagianni2018rethinking} to support consistent connectivity and high data rates.

The reviewed studies demonstrate the effective use of LP, ILP, and MILP across various resource types, highlighting their crucial role in solving complex resource allocation challenges in 5G and B5G networks. The variety of resource types considered and the successful application of these optimization techniques demonstrate the versatility and necessity of these models in modern network design and management.

\begin{table*}[ht]
\centering
\caption{Categorization of Studies based on Resource Allocation Types}
\label{Tab:Resouce_Types}
\begin{tabular}{p{0.2\textwidth}|p{0.6\textwidth}}
\hline
\textbf{Resource Allocation Problems} & \textbf{Ref.} \\
\hline
Spectrum Allocation & \cite{mounika2021downlink}, \cite{feng2017adaptive}, \cite{hadi2020patient}, \cite{swetha2018dynamic}, \cite{liu2020joint}, \cite{saeed2016control}, \cite{zhai2017adaptive}, \cite{salameh2022opportunistic}, \cite{sasikumar2021spectral}, \cite{javad2021re}, \cite{alwan2018joint}, \cite{yoon2018resource}, \cite{vlachos2017moca}, \cite{marijanovic2020multiplexing}, \cite{deb2022fog}, \cite{boutiba2023optimal}, \cite{almasaeid2023minimum}, \cite{aletri2020effect}, \cite{aletri2020optimum}, \cite{cao2021hybrid}, \cite{kokkinos2019pattern}, \cite{shokrnezhad2018joint} \\
\hline
Resource Allocation for Network Slicing & \cite{de2019optimal}, \cite{9827120}, \cite{tang2018queue}, \cite{klinkowski2022dedicated}, \cite{basu2020reverse}, \cite{jalali2020joint}, \cite{8676260}, \cite{chiaraviglio20215g}, \cite{al2019optimized}, \cite{ge2021dynamic}, \cite{ko2022pdras}, \cite{ko2021priority}, \cite{li2019latency}, \cite{boutiba2023optimal}, \cite{mei20205g}, \cite{harutyunyan2019orchestrating}, \cite{yang2019two}, \cite{fendt2018formal}, \cite{fendt2018network}, \cite{jia2017efficient}, \cite{papagianni2018rethinking}, \cite{biallach2022optimization}, \cite{mushtaq2023optimal}, \cite{gholami2023mobile}, \cite{9379507}, \cite{luu2021uncertainty}, \cite{10056963}, \cite{9685847}, \cite{gatzianas2021offline}, \cite{gang2019inter}, \cite{kokkinos2019pattern}, \cite{rayani2018slicing}, \cite{silva2023dynamic}\\
\hline
Radio Resource Management & \cite{9827120}, \cite{tang2018queue}, \cite{arouk2017multi}, \cite{mounika2021downlink}, \cite{feng2017joint}, \cite{saeed2016control}, \cite{zorello2022power}, \cite{yoon2018resource}, \cite{guo2023delay}, \cite{vlachos2017moca}, \cite{marijanovic2020multiplexing}, \cite{alsheyab2019near}, \cite{de2020optimal}, \cite{mushtaq2023optimal}, \cite{gholami2023mobile}, \cite{9929619}, \cite{cao2021hybrid} \\
\hline
Computing Resource Allocation in NFV/SDN & \cite{hasabelnaby2018joint}, \cite{sharara2023minimizing}, \cite{gkatzios2018compute}, \cite{gkatzios2019efficiency}, \cite{de2019optimal}, \cite{kazemifard2021minimum}, \cite{9827120}, \cite{tang2018queue}, \cite{mharsi2018joint}, \cite{arouk2017multi}, \cite{basu2020reverse}, \cite{xue2023joint}, \cite{xue2023joint}, \cite{zoha2017leveraging}, \cite{you2017load}, \cite{hadi2020patient}, \cite{saghezchi2017energy}, \cite{jain2021user}, \cite{zhai2017adaptive}, \cite{zorello2022power}, \cite{sharara2022policy}, \cite{guo2023delay}, \cite{fayad20235g}, \cite{deb2022fog}, \cite{zeng2019energy}, \cite{al2019optimized}, \cite{gatzianas2021energy}, \cite{lagkas2021optimized}, \cite{spinelli2022edge}, \cite{yang2023space}, \cite{gao2019deep}, \cite{yue2023delay}, \cite{jia2021toward}, \cite{prastowo2022migration}, \cite{yang2020towards}, \cite{ko2022pdras}, \cite{li2019latency}, \cite{mei20205g}, \cite{harutyunyan2019orchestrating}, \cite{yang2019two}, \cite{fendt2018formal}, \cite{fendt2018network}, \cite{papagianni2018rethinking}, \cite{de2020optimal}, \cite{de2020optimal}, \cite{morais2022placeran}, \cite{bi2019resource}, \cite{bi2019resource}, \cite{yu2019network}, \cite{mushtaq2023optimal}, \cite{gholami2023mobile}, \cite{luu2021uncertainty}, \cite{10056963}, \cite{9685847}, \cite{gatzianas2021offline}, \cite{gang2019inter}, \cite{pham2021traffic}, \cite{spinnewyn2018coordinated}, \cite{quang2018qaav}, \cite{8204056}, \cite{rayani2018slicing} \\
\hline
Energy Resource Management & \cite{hasabelnaby2018joint}, \cite{sharara2023minimizing}, \cite{gkatzios2018compute}, \cite{gkatzios2019efficiency}, \cite{ali2017joint}, \cite{fayad2022design}, \cite{kazemifard2021minimum}, \cite{klinkowski2022dedicated}, \cite{feng2017adaptive}, \cite{zoha2017leveraging}, \cite{you2017load}, \cite{monteiro2018distributed}, \cite{xie2017energy}, \cite{saghezchi2017energy}, \cite{liu2020joint}, \cite{jalali2020joint}, \cite{zhai2017adaptive}, \cite{salameh2022opportunistic}, \cite{sasikumar2021spectral}, \cite{zorello2022power}, \cite{alwan2018joint}, \cite{yoon2018resource}, \cite{guo2023delay}, \cite{deb2022fog}, \cite{zeng2019energy}, \cite{spinelli2022migration}, \cite{gatzianas2021energy}, \cite{lagkas2021optimized}, \cite{spinelli2022edge}, \cite{ge2021dynamic}, \cite{kim2022modems}, \cite{gao2019deep}, \cite{pan2020multi}, \cite{jia2017efficient}, \cite{de2020optimal}, \cite{morais2022placeran}, \cite{bi2019resource}, \cite{gatzianas2021offline}, \cite{gang2019inter}, \cite{spinnewyn2018coordinated}, \cite{cao2021hybrid}, \cite{kokkinos2019pattern}, \cite{silva2023dynamic}, \cite{shokrnezhad2018joint} \\
\hline
QoS and QoE Resource Management & \cite{hasabelnaby2018joint}, \cite{gkatzios2018compute}, \cite{ali2017joint}, \cite{tang2018queue}, \cite{mharsi2018joint}, \cite{arouk2017multi}, \cite{klinkowski2022dedicated}, \cite{mounika2021downlink}, \cite{feng2017adaptive}, \cite{feng2017joint}, \cite{zoha2017leveraging}, \cite{you2017load}, \cite{hadi2020patient}, \cite{monteiro2018distributed}, \cite{xie2017energy}, \cite{saghezchi2017energy}, \cite{jain2021user}, \cite{jalali2020joint}, \cite{salameh2022opportunistic}, \cite{zorello2022power}, \cite{sharara2022policy}, \cite{javad2021re}, \cite{alwan2018joint}, \cite{guo2023delay}, \cite{marijanovic2020multiplexing}, \cite{fayad20235g}, \cite{8676260}, \cite{al2019optimized}, \cite{spinelli2022migration}, \cite{ge2021dynamic}, \cite{kim2022modems}, \cite{yang2023space}, \cite{yue2023delay}, \cite{jia2021toward}, \cite{prastowo2022migration}, \cite{yang2020towards}, \cite{ko2022pdras}, \cite{ko2022pdras}, \cite{ko2021priority}, \cite{li2019latency}, \cite{boutiba2023optimal}, \cite{yang2019two}, \cite{fendt2018formal}, \cite{fendt2018network}, \cite{jia2017efficient}, \cite{bi2019resource}, \cite{bi2019resource}, \cite{yu2019network}, \cite{biallach2022optimization}, \cite{biallach2022optimization}, \cite{gholami2023mobile}, \cite{9379507}, \cite{luu2021uncertainty}, \cite{10056963}, \cite{9929619}, \cite{ferdosian20225g}, \cite{you2018resource}, \cite{pham2021traffic}, \cite{spinnewyn2018coordinated}, \cite{quang2018qaav}, \cite{8204056}, \cite{al2021joint}, \cite{aletri2020effect}, \cite{aletri2020optimum}, \cite{kokkinos2019pattern}, \cite{rayani2018slicing}, \cite{silva2023dynamic}, \cite{shokrnezhad2018joint} \\
\hline
Backhaul/Fronthaul Network Optimization & \cite{hasabelnaby2018joint}, \cite{gkatzios2018compute},
\cite{gkatzios2019efficiency}, \cite{de2019optimal},
\cite{ali2017joint}, \cite{fayad2022design}, \cite{kazemifard2021minimum},
\cite{arouk2017multi}, \cite{xue2023joint}, \cite{xue2023joint},
\cite{feng2017adaptive}, \cite{feng2017joint}, \cite{zoha2017leveraging},
\cite{you2017load}, \cite{monteiro2018distributed}, \cite{xie2017energy},
\cite{jain2021user}, \cite{fayad20235g}, \cite{alsheyab2019near},
\cite{zeng2019energy}, \cite{lagkas2021optimized}, \cite{yue2023delay},
\cite{li2019latency}, \cite{9379507}, \cite{kokkinos2019pattern} \\
\hline
MEC Resource Allocation & \cite{mharsi2018joint}, \cite{mounika2021downlink}, \cite{deb2022fog}, \cite{zeng2019energy}, \cite{spinelli2022migration}, \cite{gatzianas2021energy}, \cite{lagkas2021optimized}, \cite{spinelli2022edge}, \cite{ge2021dynamic}, \cite{kim2022modems}, \cite{gao2019deep}, \cite{prastowo2022migration}, \cite{pan2020multi}, \cite{harutyunyan2019orchestrating}, \cite{bi2019resource}, \cite{bi2019resource} \\
\hline
URLLC & \cite{basu2020reverse}, \cite{sharara2022policy},
\cite{marijanovic2020multiplexing},
\cite{jia2021toward},
\cite{ko2021priority},
\cite{bi2019resource},
\cite{mushtaq2023optimal},
\cite{9685847},
\cite{you2018resource} \\
\hline
Mobility Management & \cite{javad2021re}, \cite{kim2022modems}, \cite{prastowo2022migration}, \cite{yang2020towards}, \cite{papagianni2018rethinking}, \cite{cao2021hybrid} \\
\hline
\end{tabular}
\end{table*}

\subsection{Categorization of Common Objective Functions in Resource Allocation}
\label{subsec:Objective_Functions_Categories}

\begin{figure*}[t]
    \centering
    \includegraphics[scale=0.4]{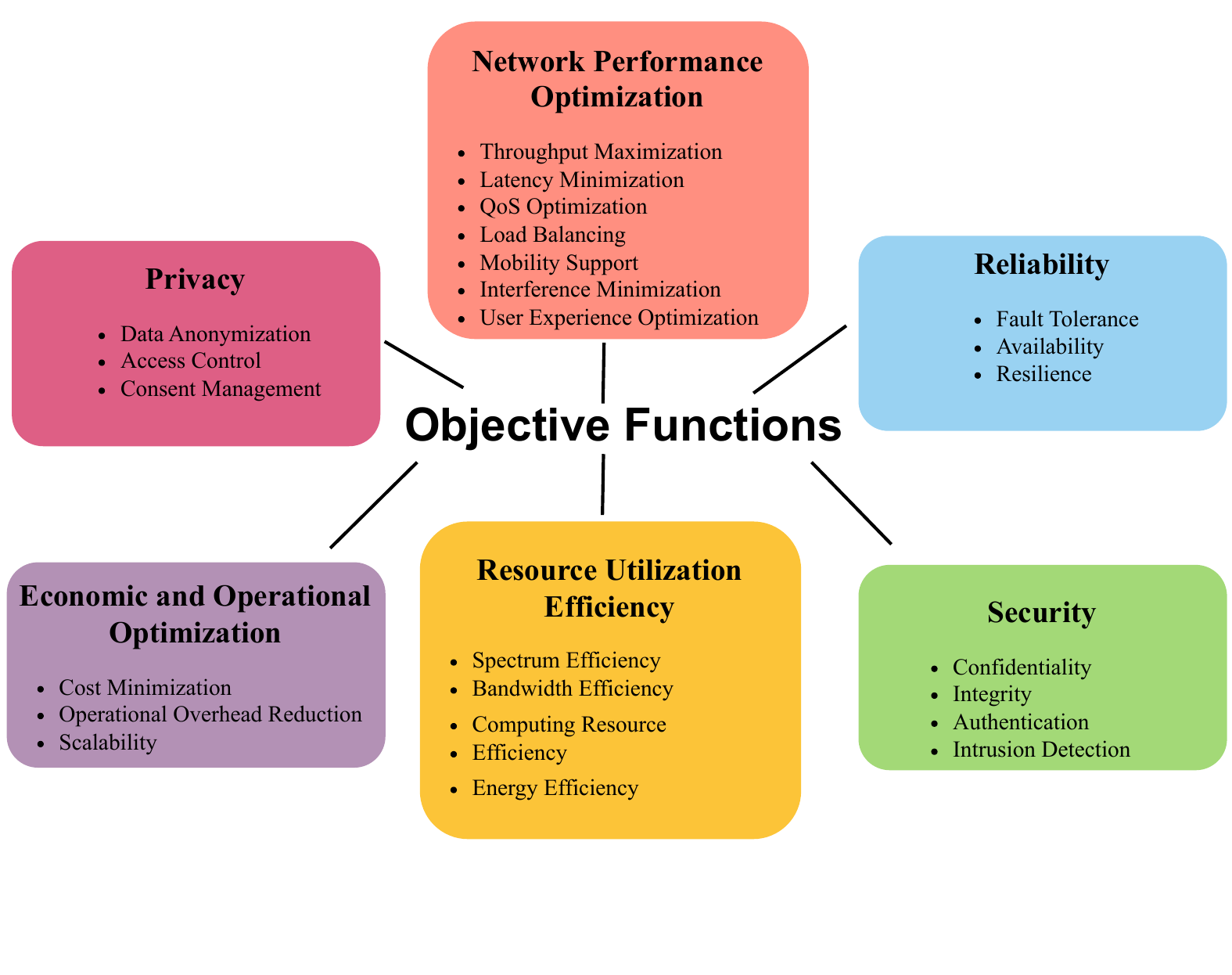}
    \caption{Objective Function Categories}
    \label{fig:objective_functions}
\end{figure*}

\begin{table*}[h]
\centering
\caption{Classification of Studies Based on Objective Function Characteristics}
\label{Tab-Obj-Fun-Cat}
\begin{tabular}{p{5cm} | p{12cm}}
\hline
\textbf{Objective Function Category} & \textbf{References} \\
\hline

Network Performance Optimization & \cite{hasabelnaby2018joint}, \cite{ali2017joint}, \cite{mounika2021downlink},  \cite{kazemifard2021minimum}, \cite{basu2020reverse},  \cite{tang2018queue}, \cite{jain2021user}, \cite{monteiro2018distributed}, \cite{jalali2020joint}, \cite{hadi2020patient}, \cite{feng2017adaptive}, \cite{feng2017joint}, \cite{you2017load}, \cite{saeed2016control}, \cite{alsheyab2019near}, \cite{marijanovic2020multiplexing}, \cite{ge2021dynamic}, \cite{kim2022modems}, \cite{salameh2022opportunistic}, \cite{guo2023delay}, \cite{yang2023space}, \cite{alwan2018joint}, \cite{vlachos2017moca}, \cite{gao2019deep}, \cite{prastowo2022migration}, \cite{ko2022pdras}, \cite{li2019latency}, \cite{boutiba2023optimal}, \cite{harutyunyan2019orchestrating}, \cite{yue2023delay}, \cite{jia2021toward}, \cite{yang2019two}, \cite{ferdosian20225g}, \cite{pan2020multi}, \cite{gatzianas2021energy}, \cite{silva2023dynamic}, \cite{al2021joint}, \cite{sharara2023minimizing}, \cite{de2019optimal},  \cite{liu2020joint}, \cite{fayad20235g}, \cite{zorello2022power}, \cite{sasikumar2021spectral},  \cite{yue2023delay}, \cite{aletri2020effect}, \cite{aletri2020optimum}, \cite{fendt2018formal}, \cite{fendt2018network},  \cite{zhai2017adaptive}, \cite{shokrnezhad2018joint}, \cite{zoha2017leveraging}, \cite{quang2018qaav},  \cite{you2018resource}, \cite{rayani2018slicing}, \cite{bi2019resource}, \cite{arouk2017multi},  \cite{9379507}, \cite{9929619}, \cite{9827120 }, \cite{8676260}, \cite{9068264}, \cite{erbati2021application}, \cite{chiaraviglio20215g}, \cite{sharara2022policy}, \cite{mharsi2018joint}, 
\cite{pham2021traffic}  \\
\hline

Resource Utilization Efficiency &  \cite{hasabelnaby2018joint}, \cite{mounika2021downlink}, \cite{tang2018queue}, \cite{basu2020reverse}, \cite{swetha2018dynamic}, \cite{alwan2018joint}, \cite{yoon2018resource}, \cite{gao2019deep}, \cite{li2019latency}, \cite{boutiba2023optimal}, \cite{harutyunyan2019orchestrating}, \cite{yue2023delay}, \cite{jia2021toward}, \cite{yang2019two}, \cite{deb2022fog}, \cite{pan2020multi}, \cite{gatzianas2021energy}, \cite{lagkas2021optimized}, \cite{silva2023dynamic}, \cite{al2021joint}, \cite{gkatzios2019efficiency}, \cite{sharara2023minimizing}, \cite{de2019optimal},  \cite{liu2020joint}, \cite{mei20205g}, \cite{sasikumar2021spectral},  \cite{yue2023delay}, \cite{kokkinos2019pattern}, \cite{fendt2018formal}, \cite{gkatzios2018compute},  \cite{saghezchi2017energy}, \cite{xie2017energy}, \cite{jia2017efficient}, \cite{shokrnezhad2018joint}, \cite{zoha2017leveraging},  \cite{spinnewyn2018coordinated}, \cite{zeng2019energy}, \cite{de2020optimal}, \cite{al2019optimized}, \cite{morais2022placeran}, \cite{arouk2017multi}, \cite{gholami2023mobile}, \cite{9929619}, \cite{spinelli2022migration}, \cite{spinelli2022edge}, \cite{gang2019inter}, \cite{mharsi2018joint}, 
\cite{pham2021traffic}   \\
\hline

Economic and Operational Optimization & \cite{alsheyab2019near}, \cite{marijanovic2020multiplexing}, \cite{ge2021dynamic}, \cite{prastowo2022migration}, \cite{harutyunyan2019orchestrating}, \cite{silva2023dynamic}, \cite{xue2023joint}, \cite{jia2017efficient}, \cite{quang2018qaav},  \cite{spinnewyn2018coordinated}, \cite{rayani2018slicing}, \cite{papagianni2018rethinking}, \cite{mushtaq2023optimal}, \cite{yu2019network}, \cite{biallach2022optimization}, \cite{gholami2023mobile}, \cite{luu2021uncertainty}, \cite{10056963}, \cite{9827120}, \cite{9685847}, \cite{9068264}, \cite{spinelli2022migration}, \cite{spinelli2022edge}, \cite{chiaraviglio20215g}, \cite{klinkowski2022dedicated}, \cite{gang2019inter}, \cite{cao2021hybrid}, \cite{fayad2022design}, \cite{yang2020towards}, \cite{8204056}  \\
\hline

Reliability  &  \cite{hasabelnaby2018joint},  \cite{tang2018queue}, \cite{ferdosian20225g}, \cite{javad2021re}, \cite{mushtaq2023optimal}, \cite{8676260}, \cite{erbati2021application}  \\
\hline


\end{tabular}
\end{table*}


In the context of optimizing various aspects of 5G and beyond network architectures, LP, ILP, and MILP have been instrumental in addressing a wide range of objective functions. To understand the application scope of these mathematical models, we categorized the selected studies into four primary objective function categories: Network Performance Optimization, Resource Utilization Efficiency, Economic and Operational Optimization, and Reliability. Table \ref{Tab-Obj-Fun-Cat} organizes the selected studies based on their objective functions, making it easier to see how each study approaches resource allocation. 
\noindent

Network Performance Optimization encompasses the fundamental goals of throughput maximization, latency minimization, QoS optimization, load balancing, mobility support, and interference minimization \cite{jain2021user, ko2022pdras, li2019latency}. In 5G and B5G networks, resource allocation algorithms generally aim to minimize latency, maximize throughput, and ensure reliable communication to meet stringent performance requirements.



The resource utilization efficiency objective aims to optimize the utilization of network resources, including spectrum, bandwidth, and computing resources, to maximize efficiency and minimize waste \cite{boutiba2023optimal, yang2019two, pan2020multi}. In 5G and B5G networks, where network densification and massive MIMO technologies are employed to meet increasing demand, efficient resource allocation is essential to maximize limited resources. 


The objective of economic and operational optimization focuses on minimizing costs and operational overhead while maintaining the desired level of service quality. In 5G and B5G networks, where infrastructure deployment and complex networks' operation need significant investments, optimizing economic and operational aspects is critical \cite{jia2017efficient, morais2022placeran, gang2019inter}. Cost-effective deployment strategies, energy-efficient network design, and automated network management are essential to reduce capital and operational expenses.


The reliability objective aims to ensure the reliability of network operations and data transmission. In 5G and B5G networks, where a wide range of critical applications and services are supported, ensuring reliability is paramount to protect against various threats and vulnerabilities \cite{hasabelnaby2018joint, tang2018queue, ferdosian20225g}.


\subsection{Constraint Modeling and Types}
\label{sec:Constraints}


\begin{figure*}[htbp]
    \centering
     \includegraphics[width=0.7\linewidth]{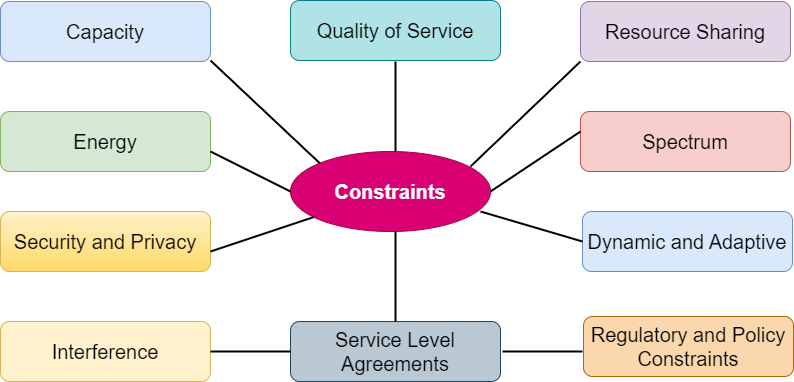}
    \caption{Categorization of Constraints}
    \label{fig:constraints}
\end{figure*}

This section presents a categorized review of the constraints utilized in optimization functions modelled as LP, ILP, and MILP for 5G and B5G networks. By grouping related studies, we focus on key constraint types such as capacity, QoS, energy efficiency, and interference, among others, highlighting their importance in efficient network resource allocation and optimization. Figure \ref{fig:constraints} shows the categories and Table \ref{Tab-Constraints-Cat} categorizes the selected 103 studies based on various constraints relevant to resource allocation in B5G networks. 


Capacity constraints ensure optimal resource allocation to meet bandwidth, user, and service requirements in 5G and B5G networks. Common themes in the selected studies include preventing overload and maximizing efficiency. Studies such as \cite{mushtaq2023optimal} and \cite{biallach2022optimization} address server utilization and computational resource limits, while \cite{gholami2023mobile} and \cite{9379507} focus on optimal resource allocation at network nodes and slices. 


QoS constraints guarantee network performance by addressing latency, throughput, and reliability. Studies like \cite{marijanovic2020multiplexing} and \cite{silva2023dynamic} focus on minimizing latency and maintaining data rates, while \cite{ge2021dynamic} and \cite{spinnewyn2018coordinated} dynamically manage resources to meet latency and bandwidth demands. These constraints are essential for ensuring performance across diverse applications, such as video streaming and real-time communication.

Energy efficiency is a key concern, especially with the growing number of devices and data-hungry applications. \cite{sharara2023minimizing} and \cite{saghezchi2017energy} emphasize keeping energy consumption within limits for transmission and device operations. \cite{xie2017energy} and \cite{zoha2017leveraging} explore strategies for reducing energy consumption in caching and scheduling, while maintaining performance. 


Efficient spectrum management is essential for supporting high network capacity and avoiding interference. \cite{sasikumar2021spectral}, \cite{gholami2023mobile}, and \cite{silva2023dynamic} address spectrum constraints by optimizing resource blocks and adhering to allocation policies. 

Security and privacy constraints address the challenges of protecting data and ensuring network integrity. \cite{gholami2023mobile} and \cite{luu2021uncertainty} tackle demand uncertainty and resource provisioning to maintain security, while \cite{cao2021hybrid} explores Quantum Key Distribution (QKD) for securing optical networks. These works emphasize the importance of securing data across expanding and complex networks.

Dynamic and adaptive constraints enable networks to adjust in real time to changing conditions. \cite{salameh2022opportunistic}, \cite{boutiba2023optimal}, and \cite{guo2023delay} highlight the importance of adapting resource management and scheduling algorithms to ensure reliable performance. These studies showcase how adaptive approaches help maintain network efficiency in response to fluctuating traffic and user demands.

Managing interference is critical, especially in densely deployed 5G networks which is ensured by interference constraints. \cite{mounika2021downlink} and \cite{al2021joint} address interference in beamforming and device-to-device communication, while \cite{lagkas2021optimized} and \cite{sharara2023minimizing} propose strategies for reducing interference in HetNets and dense environments. These constraints ensure that network resources are efficiently allocated without compromising performance due to interference.

Service Level Agreements (SLA) constraints ensure networks meet contractual obligations for latency, throughput, and reliability. \cite{biallach2022optimization}, \cite{marijanovic2020multiplexing}, and \cite{mharsi2018joint} address how resource allocation strategies help meet SLA requirements, particularly in network slicing and URLLC.

Regulatory constraints ensure networks comply with government and international standards. \cite{chiaraviglio20215g}, \cite{yue2023delay}, and \cite{hasabelnaby2018joint} address spectrum management, data privacy, and environmental standards. These works highlight the impact of regulations on network deployment and resource allocation strategies, especially concerning spectrum use and energy consumption.

\begin{table*}[ht]
\centering
\caption{Classification of Studies Based on Constraints}
\label{Tab-Constraints-Cat}
\begin{tabular}{p{0.2\textwidth}|p{0.6\textwidth}}
\hline

\textbf{Constraints Category} & \textbf{References} \\

\hline
Quality-of-Service (QoS) Constraints & \cite{hasabelnaby2018joint}, \cite{sharara2023minimizing}, \cite{gkatzios2019efficiency}, \cite{ali2017joint}, \cite{feng2017adaptive}, \cite{feng2017joint}, \cite{zoha2017leveraging}, \cite{you2017load}, \cite{zhai2017adaptive}, \cite{zorello2022power}, \cite{sharara2022policy}, \cite{javad2021re}, \cite{alwan2018joint}, \cite{yoon2018resource}, \cite{vlachos2017moca}, \cite{marijanovic2020multiplexing}, \cite{fayad20235g}, \cite{deb2022fog}, \cite{8676260}, \cite{alsheyab2019near}, \cite{gatzianas2021energy}, \cite{lagkas2021optimized}, \cite{spinelli2022edge}, \cite{ge2021dynamic}, \cite{yang2023space}, \cite{gao2019deep}, \cite{yue2023delay}, \cite{jia2021toward}, \cite{prastowo2022migration}, \cite{ko2022pdras}, \cite{ko2021priority}, \cite{li2019latency}, \cite{boutiba2023optimal}, \cite{mei20205g}, \cite{harutyunyan2019orchestrating}, \cite{yang2019two}, \cite{fendt2018formal}, \cite{fendt2018network}, \cite{gholami2023mobile}, \cite{9379507}, \cite{10056963}, \cite{9929619}, \cite{ferdosian20225g}, \cite{erbati2021application}, \cite{pham2021traffic}, \cite{spinnewyn2018coordinated}, \cite{quang2018qaav}, \cite{almasaeid2023minimum}, \cite{al2021joint}, \cite{silva2023dynamic}, \cite{9068264}\\

\hline

Energy Efficiency Constraints & \cite{hasabelnaby2018joint}, \cite{sharara2023minimizing}, \cite{gkatzios2018compute}, \cite{gkatzios2019efficiency}, \cite{fayad2022design}, \cite{basu2020reverse}, \cite{zoha2017leveraging}, \cite{xie2017energy}, \cite{saghezchi2017energy}, \cite{liu2020joint}, \cite{zhai2017adaptive}, \cite{sasikumar2021spectral}, \cite{zorello2022power}, \cite{fayad20235g}, \cite{deb2022fog}, \cite{zeng2019energy}, \cite{spinelli2022migration}, \cite{gatzianas2021energy}, \cite{lagkas2021optimized}, \cite{spinelli2022edge}, \cite{ge2021dynamic}, \cite{pan2020multi}, \cite{yang2019two}, \cite{jia2017efficient}, \cite{gatzianas2021offline}, \cite{gang2019inter}, \cite{spinnewyn2018coordinated} \\ 

\hline

Spectrum Constraints & \cite{hasabelnaby2018joint}, \cite{sharara2023minimizing}, \cite{ali2017joint}, \cite{mounika2021downlink}, \cite{feng2017joint}, \cite{swetha2018dynamic}, \cite{liu2020joint}, \cite{saeed2016control}, \cite{sasikumar2021spectral}, \cite{javad2021re}, \cite{alwan2018joint}, \cite{yoon2018resource}, \cite{vlachos2017moca}, \cite{deb2022fog}, \cite{8676260}, \cite{alsheyab2019near}, \cite{zeng2019energy}, \cite{al2019optimized}, \cite{lagkas2021optimized}, \cite{papagianni2018rethinking}, \cite{morais2022placeran}, \cite{yu2019network}, \cite{gholami2023mobile}, \cite{9685847}, \cite{almasaeid2023minimum}, \cite{aletri2020effect}, \cite{aletri2020optimum}, \cite{kokkinos2019pattern}, \cite{silva2023dynamic} \\ 

\hline

Security and Privacy Constraints & \cite{jalali2020joint}, \cite{8676260}, \cite{gholami2023mobile}, \cite{luu2021uncertainty}, \cite{cao2021hybrid} \\ \hline

Dynamic and Adaptive Constraints & \cite{pham2021traffic}, \cite{kokkinos2019pattern}, \cite{zhai2017adaptive}, \cite{feng2017adaptive}, \cite{swetha2018dynamic}, \cite{8204056}, \cite{yang2019two}, \cite{saghezchi2017energy}, \cite{xie2017energy}, \cite{feng2017joint}, \cite{tang2018queue}, \cite{alwan2018joint}, \cite{hasabelnaby2018joint}, \cite{ali2017joint}, \cite{zoha2017leveraging}, \cite{aletri2020optimum}, \cite{jia2021toward}, \cite{guo2023delay}, \cite{boutiba2023optimal}, \cite{harutyunyan2019orchestrating}, \cite{morais2022placeran}, \cite{de2020optimal}, \cite{you2018resource}, \cite{vlachos2017moca}, \cite{kazemifard2021minimum}, \cite{gang2019inter}, \cite{gao2019deep}, \cite{saeed2016control}, \cite{luu2021uncertainty}, \cite{basu2020reverse}, \cite{sharara2022policy}, \cite{yang2020towards}, \cite{quang2018qaav}, \cite{deb2022fog}, \cite{de2019optimal}, \cite{biallach2022optimization}, \cite{10056963}, \cite{mharsi2018joint}, \cite{prastowo2022migration}, \cite{spinelli2022migration}, \cite{arouk2017multi}, \cite{klinkowski2022dedicated}, \cite{li2019latency}, \cite{mei20205g}, \cite{gholami2023mobile}, \cite{salameh2022opportunistic}, \cite{yoon2018resource}, \cite{monteiro2018distributed}, \cite{fendt2018network}, \cite{hadi2020patient}, \cite{marijanovic2020multiplexing}, \cite{aletri2020effect}\\ \hline

Interference Constraints & \cite{hasabelnaby2018joint}, \cite{sharara2023minimizing}, \cite{ali2017joint}, \cite{mharsi2018joint}, \cite{mounika2021downlink}, \cite{xue2023joint}, \cite{feng2017joint}, \cite{zoha2017leveraging}, \cite{you2017load}, \cite{jain2021user}, \cite{saeed2016control}, \cite{jalali2020joint}, \cite{salameh2022opportunistic}', \cite{sasikumar2021spectral}, \cite{javad2021re}, \cite{alwan2018joint}, \cite{yoon2018resource}, \cite{vlachos2017moca}, \cite{deb2022fog}, \cite{alsheyab2019near}, \cite{lagkas2021optimized}, \cite{boutiba2023optimal}, \cite{yang2019two}, \cite{al2021joint}] \\ \hline

SLAs Constraints & \cite{tang2018queue}, \cite{arouk2017multi}, \cite{mharsi2018joint},  \cite{marijanovic2020multiplexing}, \cite{biallach2022optimization} \\ \hline
Regulatory and Policy Constraints & \cite{hasabelnaby2018joint}, \cite{chiaraviglio20215g}, \cite{yue2023delay}, \cite{erbati2021application},  \\ \hline

    \end{tabular}
   
\end{table*}

\section{Solution Techniques and Approaches}
\label{Solution_Techniques}

This section discusses solution techniques and approaches employed by the selected studies for solving the optimization problems modelled as LP, ILP, and MILP. Table \ref{Tab-Problem_Modelling} categorizes the selected studies based on their solution types. 

\subsection{Linear Programming based Modelling and Solution}

LP has long been established as a powerful mathematical tool for optimizing decision-making processes in various domains. In the context of 5G and B5G networks, LP has gained prominence due to its ability to model complex resource allocation, scheduling, and network planning problems, focusing on achieving optimal solutions while ensuring constraints are met. This section presents a detailed review of the papers that employed LP techniques to address optimization challenges in 5G and B5G networks.


\subsubsection{Direct Formulation as LP}
Several studies have directly formulated their problems as LP models. For instance, Saghezchi et al. \cite{saghezchi2017energy} tackled the relay selection problem, aiming to optimize the assignment of relays to sources to maximize energy savings. They noted that due to the inherent structure of the problem, the LP relaxation naturally produced integer solutions, thus bypassing the need for explicit integer constraints. 

Guo et al. \cite{guo2023delay} focused on minimizing the probability of transmission delays exceeding a certain threshold, while adhering to reliability and bandwidth constraints. They modeled their problem as an LP, employing variables that represent system states and actions. Their solution provides an optimal strategy for encoding and transmitting data, effectively balancing delay and reliability. 

Yang et al. \cite{yang2019two} applied LP to optimize radio resource allocation within edge cloud slices, demonstrating LP’s capability to enhance resource allocation efficiency in a specific network segment. 

\subsubsection{Transformation to LP}

In addition to the direct applications, several studies transformed their non-linear or complex optimization problems into LP problems. This approach demonstrates how LP can be integrated into a broader framework to handle complex scenarios. Feng et al. \cite{feng2017adaptive} faced a non-linear optimization problem that they decomposed into several linear sub-problems, leveraging LP techniques to address parts of the problem.  Jalali et al. \cite{jalali2020joint} converted a problem involving the minimization of the Euclidean distance between received and estimated transmitted signals into an LP problem by redefining the objective function and constraints. 

Sasikumar et al. \cite{sasikumar2021spectral} initially formulated their optimization problem as a multi-objective optimization problem (MOO), which was inherently a mixed-integer non-linear programming (MINLP) problem. They employed the epsilon-constraint and linearization techniques to convert it into a single-objective mixed-integer linear optimization problem. This reformulation enabled the use of LP approaches. Yue et al. \cite{yue2023delay} addressed an optimization problem involving the placement of VNFs while ensuring resource constraints and delay requirements. They modelled this as a weighted graph-matching problem, which was then converted into an LP problem to find the optimal mapping between VNFs and physical nodes. 

Ko et al. \cite{ko2022pdras, ko2021priority} worked on maximizing the average prioritized QoS across all network slices while adhering to resource limits. They formulated this as a Constrained Markov Decision Process (CMDP) and converted it into an equivalent LP problem for efficient solving. Lastly,Singh and Murthy \cite{8204056} tackled a problem fundamentally framed as an integer program but utilized LP relaxation to allow fractional block assignments. This relaxation provided utility values that guided block allocation in the more complex integer program.

\subsubsection{Limitations of LP in 5G and B5G Network Optimization}

In 5G and B5G networks, LP is useful for various optimization problems but has limitations due to its simplicity. LP often requires problem decomposition or approximation, leading to approximate rather than exact solutions. For complex issues like interference management and network slicing, more advanced models like integer programming, non-linear optimization, or dynamic programming are needed. Thus, while LP is efficient, it may not fully address intricate network challenges.




\begin{table*}[ht]
\centering
\caption{Summary of Problem Modelling Approaches and Solution Techniques}
\label{Tab-Problem_Modelling}
\begin{tabular}{p{3cm}|p{4cm}|p{8cm}}
\hline
\textbf{Modelling Approach} & \textbf{Solution Technique} & \textbf{Ref}\\

\hline
LP  & -  & \cite{feng2017adaptive}, \cite{saghezchi2017energy}, \cite{jalali2020joint}, \cite{sasikumar2021spectral}, \cite{guo2023delay}, \cite{yue2023delay}, \cite{ko2022pdras}, \cite{ko2021priority}, \cite{yang2019two}, \cite{you2018resource}, \cite{8204056}\\
\hline

ILP & No Heuristic (ILP Solver) & \cite{hasabelnaby2018joint}, \cite{de2019optimal}, \cite{klinkowski2022dedicated}, \cite{swetha2018dynamic}, \cite{pan2020multi}, \cite{fendt2018network}, \cite{mushtaq2023optimal},  \cite{biallach2022optimization}, \cite{morais2022placeran}, \cite{rayani2018slicing} \\
\hline

MILP & No Heuristic (MILP Solver) & \cite{ali2017joint}, \cite{8676260}, \cite{zeng2019energy}, \cite{gao2019deep}, \cite{prastowo2022migration}, \cite{fendt2018formal}, \cite{papagianni2018rethinking}, \cite{9685847}, \cite{aletri2020effect}, \cite{aletri2020optimum}, \cite{silva2023dynamic}, \cite{kazemifard2021minimum}, \cite{monteiro2018distributed} \\
\hline


ILP  & Other Solutions & \cite{gkatzios2018compute}, \cite{gkatzios2019efficiency}, \cite{fayad2022design}, \cite{9827120}, \cite{mharsi2018joint}, \cite{arouk2017multi}, \cite{mounika2021downlink}, \cite{xue2023joint} [32], \cite{feng2017joint}, \cite{zoha2017leveraging}, \cite{xie2017energy}, \cite{marijanovic2020multiplexing}, \cite{fayad20235g}, \cite{sharara2022policy}, \cite{javad2021re}, \cite{alwan2018joint}, \cite{vlachos2017moca}, \cite{alsheyab2019near}, \cite{spinelli2022migration}, \cite{lagkas2021optimized}, \cite{ge2021dynamic}, \cite{kim2022modems}, \cite{yang2020towards}, \cite{saeed2016control}, \cite{spinelli2022edge}, \cite{li2019latency}, \cite{mei20205g}, \cite{jia2017efficient}, \cite{9379507}, \cite{10056963}, \cite{ferdosian20225g}, \cite{erbati2021application}, \cite{spinnewyn2018coordinated}, \cite{quang2018qaav}, \cite{almasaeid2023minimum}, \cite{cao2021hybrid}, \cite{kokkinos2019pattern}, \cite{9068264} \\

\hline

MILP & Other Solutions & \cite{sharara2023minimizing}, \cite{basu2020reverse}, \cite{hadi2020patient}, \cite{you2017load}, \cite{liu2020joint}, \cite{jain2021user}, \cite{zorello2022power}, \cite{zhai2017adaptive}, \cite{yoon2018resource}, \cite{deb2022fog}, \cite{chiaraviglio20215g}, \cite{al2019optimized}, \cite{chu2023hirn}, \cite{gatzianas2021energy}, \cite{yang2023space}, \cite{jia2021toward}, \cite{boutiba2023optimal}, \cite{harutyunyan2019orchestrating}, \cite{bi2019resource}, \cite{de2020optimal}, \cite{gholami2023mobile}, \cite{yu2019network}, \cite{luu2021uncertainty}, \cite{9929619}, \cite{gatzianas2021offline}, \cite{gang2019inter}, \cite{pham2021traffic}, \cite{al2021joint}, \cite{shokrnezhad2018joint}  \\






\hline

\hline

\end{tabular}
  
\end{table*}

\subsection{ILP and MILP without heuristic}
This subsection focuses on studies that model network resource allocation problems using ILP or MILP without employing or proposing any heuristics. These optimization models are frequently used due to their ability to handle discrete variables and capture complex constraints. However, ILP and MILP problems are known to be NP-hard, meaning that finding an optimal solution is computationally challenging and often infeasible for large-scale networks. Despite these difficulties, several studies have adopted these models to provide exact solutions. Problem sizes are often kept small to enable ILP solutions within reasonable timeframes. 

De et al. \cite{de2019optimal} addressed VNF deployment and resource allocation in hybrid cloud infrastructures, while Klinkowski et al. \cite{klinkowski2022dedicated} optimized wavelength placement and network paths for 5G survivability using ILP with the CPLEX solver. Swetha et al. \cite{swetha2018dynamic} tackled dynamic channel allocation using a prediction-based model with ILP, and Pan et al. \cite{pan2020multi} applied linearization methods for efficient task offloading to mobile devices. Fendt et al. \cite{fendt2018network} solved moderately large ILP instances, and Mushtaq et al. \cite{mushtaq2023optimal} demonstrated feasibility with the Gurobi optimizer.

Hasabelnaby et al. \cite{hasabelnaby2018joint} modelled the problem as an ILP and used an exhaustive search to find the optimal solution. To reduce the computational cost, they proposed storing optimal configurations in a lookup table indexed by loss intervals. While this improves efficiency for real-time decisions, it sacrifices some optimality as the table may not cover all scenarios.


Similarly, several studies have employed MILP formulations to address various optimization problems in network systems. These include joint optimization of user association, admission control, and power allocation in H-CRANs \cite{ali2017joint}, resource allocation in 5G core network slicing \cite{8676260, 9685847}, energy efficiency in 5G RANs \cite{zeng2019energy}, optimal task partitioning and offloading decisions in MEC environments \cite{gao2019deep}, and computation and communication resource allocation in B5G edge networks \cite{prastowo2022migration}. Additionally, MILP has been applied to network slice embedding \cite{fendt2018formal}, minimizing resource provisioning costs for LTE operators \cite{papagianni2018rethinking}, resource allocation in optical wireless communication systems \cite{aletri2020effect, aletri2020optimum}, and minimizing the number of UAV-deployed gateways for network coverage \cite{silva2023dynamic}.




ILP and MILP guarantee optimal solutions within defined constraints, ensuring the best possible outcomes according to established criteria. This precision is beneficial as it avoids the approximation errors common in heuristic methods. Many studies using ILP and MILP rely on solvers like Gurobi or CPLEX, known for their efficiency in managing complex models. Despite their NP-hard nature, these solvers have shown remarkable scalability, particularly with advances in computational power and optimization algorithms, allowing them to handle large-scale problems efficiently.

However, solving ILP and MILP problems can be computationally intensive for large instances, demanding substantial resources and time. Although solver efficiency has improved, complex models can still be time-consuming, especially with large-scale or highly constrained problems. In some cases, particularly with extremely large or complex problems, ILP and MILP might not be practical due to computational limitations. Alternative approaches like metaheuristics or approximation algorithms may offer more feasible solutions for obtaining near-optimal results within reasonable timeframes.

\begin{table*}
\centering
\caption{Categorization based on Algorithms Types (ILP and MILP)}
\label{tab:ILP-MILP-Heuristics}
\begin{tabular}{p{0.30\linewidth}|p{0.05\linewidth}|p{0.53\linewidth}}
\hline
\textbf{Solution Type} & \textbf{Model} & \textbf{Reference} \\
\hline

\hline
\multirow{2}{*}{Exact Algorithms with Heuristic Components} & ILP & \cite{mounika2021downlink}, \cite{javad2021re}, \cite{saeed2016control}, \cite{lagkas2021optimized}, \cite{li2019latency}, \cite{spinnewyn2018coordinated}, \cite{alwan2018joint}, \cite{9379507}, \cite{de2020optimal}, \cite{erbati2021application}, \cite{kokkinos2019pattern}\\
\cline{2-3}
 & MILP & \cite{you2017load}, \cite{hadi2020patient}, \cite{jain2021user}, \cite{al2019optimized}, \cite{chu2023hirn}, \cite{deb2022fog}, \cite{luu2021uncertainty}, \cite{gholami2023mobile}\\
\hline

\multirow{2}{*}{MetaHeuristics} & ILP &  \cite{fayad2022design}, \cite{xie2017energy}, \cite{ge2021dynamic}, \cite{jia2017efficient}\\
\cline{2-3}
 & MILP & - \\

\hline
\multirow{2}{*}{Decomposition-Based Heuristics} & ILP & \cite{gkatzios2019efficiency}, \cite{marijanovic2020multiplexing}, \cite{yang2020towards}, \cite{jia2017efficient}, \cite{shokrnezhad2018joint} \\
\cline{2-3}
 & MILP & \cite{gkatzios2018compute}, \cite{yu2019network} \\
\hline
\multirow{2}{*}{Local Search and Constructive Heuristics} & ILP & \cite{9827120}, \cite{arouk2017multi}, \cite{vlachos2017moca}, \cite{xue2023joint}, \cite{fayad20235g}, \cite{spinelli2022migration}, \cite{kim2022modems}, \cite{mei20205g}, \cite{mei20205g}, \cite{10056963}, \cite{cao2021hybrid}
 \\
\cline{2-3}
 & MILP &  \cite{tang2018queue}, \cite{basu2020reverse}, \cite{liu2020joint}, \cite{zorello2022power}, \cite{yoon2018resource}, \cite{gatzianas2021energy}, \cite{spinelli2022edge}, \cite{harutyunyan2019orchestrating}, \cite{gang2019inter}, \cite{gatzianas2021offline}, \cite{al2021joint}  \\
 
\hline
\multirow{2}{*}{Problem Specific Heuristics} & ILP & \cite{zoha2017leveraging}, \cite{zhai2017adaptive}, \cite{alsheyab2019near},\\
\cline{2-3}
 & MILP &  \cite{chiaraviglio20215g}, \cite{ferdosian20225g} \\
\hline
\multirow{2}{*}{Hybrid Heuristics and Reinforcement Learning} & ILP &  \cite{sharara2022policy}, \cite{9068264} \\
\cline{2-3}
 & MILP & \cite{boutiba2023optimal}, \cite{pham2021traffic} \\
\hline
\multirow{2}{*}{Heuristic based Relaxations} & ILP & - \\
\cline{2-3}
 & MILP & \cite{9929619} \\
\hline
\multirow{2}{*}{Approximation Algorithms} & ILP & \cite{sharara2023minimizing}, \cite{mharsi2018joint}, \cite{feng2017joint}, \cite{salameh2022opportunistic}, \cite{almasaeid2023minimum}
 \\
\cline{2-3}
 & MILP & \cite{yang2023space}, \cite{bi2019resource}, \cite{quang2018qaav}\\
\hline
\end{tabular}
\end{table*}

\subsection{ILP and MILP with other algorithms}
The inherent complexity of ILP and MILP problems often necessitates exploring alternative solution techniques beyond exact methods. This section explores a variety of heuristic and approximation algorithms that offer practical trade-offs between solution quality and computational efficiency. Table \ref{tab:ILP-MILP-Heuristics} categorizes those selected studies which used approximation algorithms for obtaining a sub-optimal solution of ILP and MILP optimization problems.   


\subsubsection{Exact Algorithms with Heuristic Components}
Exact algorithms with heuristic components aim to find the optimal solution by systematically exploring the solution space, incorporating heuristic strategies to expedite the search and avoid local optima. 
The selected studies can be further categorized into the following subcategories:

\begin{itemize}
    \item \textbf{Prioritization and Selection:} The core principle of these heuristics is to strategically prioritize or select a subset of variables or constraints, thereby reducing the problem's complexity. This can involve prioritizing users based on specific metrics, focusing on specific network elements, or pre-filtering variables to streamline the optimization process. For instance, Mounika et. al \cite{mounika2021downlink} prioritized users based on a weighted rate metric and selected a subset for optimization, while Lagkas et. al \cite{lagkas2021optimized} prioritized Remote Radio Heads (RRHs) based on coverage to guide resource allocation. Alwan et al. \cite{alwan2018joint} pre-filtered flows to reduce the problem size before applying the Branch-and-Cut method. The heuristic proposed by Deb et al. \cite{de2020optimal} prioritized VNF chain requests based on computational requirements and used a best-fit approach for cloud selection.

    \item \textbf{Greedy Algorithms: }These heuristics iteratively construct solutions by making locally optimal choices at each step. They prioritize immediate gains, aiming for good solutions in a reasonable timeframe, although they might not guarantee global optimality. For instance, the Greedy Chain Selection (GCS) algorithm by Kokkinos et. al \cite{spinnewyn2018coordinated} selected the chain that provided the most benefit relative to its cost.

    \item \textbf{Branch-and-Bound with Enhancements:} The branch-and-bound algorithm is a fundamental approach for solving ILPs and MILPs. The surveyed literature showcases enhancements to this algorithm, often incorporating problem-specific heuristics to accelerate the search for optimal or near-optimal solutions. Li et al. \cite{li2019latency} used branch-and-bound for network slicing resource allocation.


    \item \textbf{Iterative Refinement: }The essence of these heuristics lies in iteratively improving an initial solution. They strategically adjust variables or constraints based on predefined criteria to converge toward an optimal or near-optimal solution. For instance, You et al. \cite{you2017load} introduced the MinL algorithm, which iteratively refines cell-UE association to minimize load on the network.

    \item \textbf{Traffic-Aware Placement: }These heuristics consider traffic patterns and network loads to guide the placement of resources or network elements. By strategically positioning elements based on traffic demands, they aim to optimize network performance and resource utilization. Al et al. \cite{al2019optimized} proposed two heuristics that strategically place virtual machines in 5G network function virtualization to minimize power consumption due to traffic routing.
\end{itemize}


\subsubsection{Metaheuristics}
Metaheuristics provide general strategies for exploring the solution space and escaping local optima. These high-level frameworks, inspired by natural phenomena, include Genetic Algorithms (which mimic evolution), Simulated Annealing (which simulates cooling metals), Tabu Search (which uses memory to restrict certain moves), Particle Swarm Optimization (which models social behaviour), Ant Colony Optimization (which imitates ant foraging), and Memetic Algorithms (which combine local search with metaheuristics). These methods can handle large-scale and complex problems, often providing good solutions in reasonable time, and are flexible and adaptable. However, they do not guarantee optimality, and their performance can be sensitive to parameter settings.

Jia et al. \cite{jia2017efficient} investigated the use of Chemical Reaction Optimization (CRO) to tackle the ILP problem of caching resource allocation in 5G networks. The CRO algorithm, inspired by chemical reactions, offered flexibility, global search capability, and computational efficiency, making it suitable for complex ILP problems. However, it required careful parameter tuning and lacked theoretical guarantees of optimality. Xie et al. \cite{xie2017energy} proposed a Quantum-inspired Evolutionary Algorithm (QEA) to optimize energy efficiency in heterogeneous cellular networks. The QEA leveraged quantum computing principles to explore a larger solution space and potentially find the global optimum. While it demonstrated superior performance compared to classical algorithms, QEA could be computationally complex and sensitive to parameter choices. Fayad et al. \cite{fayad2022design} employed K-means clustering and a Genetic Algorithm to address the scalability challenges of the ILP problem for cost-effective optical fronthaul deployment.

\subsubsection{Local Search and Constructive Heuristics}
Local search methods, like Greedy Algorithms that make the best choice at each step, Hill Climbing, which seeks better solutions until it reaches a local optimum, and Variable Neighborhood Search, which explores different areas, all aim to improve existing solutions. Constructive Heuristics, such as GRASP, which combines greedy construction with local search, Sequential Heuristics, which build solutions step-by-step, and Relax-and-Fix, which decomposes the problem and fixes some variables while optimizing others, focus on building solutions from scratch. These methods are simple to implement, often provide good solutions quickly, and can be easily customized, but they can get trapped in local optima and may not explore the solution space thoroughly. 

The selected studies that use local search and constructive heuristics can be further divided into the following categories:

\begin{itemize}
    \item \textbf{Greedy Heuristics:}  Several papers employ greedy heuristics to construct solutions incrementally. For instance, \cite{tang2018queue} utilizes a greedy algorithm for VNF placement, prioritizing nodes that maximize a utility function. Similarly, \cite{gatzianas2021energy} and \cite{harutyunyan2019orchestrating} adopt greedy strategies for path selection and component placement, respectively. The advantage of these methods lies in their computational efficiency, making them suitable for real-time or large-scale scenarios.

    \item \textbf{Hybrid Approaches:} A few papers combine constructive and local search elements. Zorello et al. \cite{zorello2022power} constructed an initial solution using centrality-based placement and then refined it with local search to minimize power consumption. Gatzianas et al. \cite{gatzianas2021offline} built a solution step-by-step but explores alternative paths if constraints are violated, showcasing a blend of both strategies. 

    \item \textbf{Problem-Specific Local Search-based Heuristics:} Some papers propose heuristics tailored to the specific problem at hand. \cite{basu2020reverse} introduces the RPFM to minimize H-plane traffic in SDN networks. \cite{yoon2018resource} employs a three-step heuristic for resource allocation in NOMA-based D2D communication. These problem-specific heuristics leverage domain knowledge to achieve effective solutions, albeit potentially at the cost of generalizability.

    \item \textbf{Decomposition and Transformation with local search} Some papers decompose complex problems into smaller, more manageable sub-problems. Liu et al. \cite{liu2020joint} transformed a stochastic optimization problem into a MILP and then relaxed it into an LP under specific conditions, enabling efficient solving. Al-Wesabi \cite{al2021joint} decomposed the main problem into access control and time-domain scheduling sub-problems, each addressed with tailored heuristics. This approach allowed for a more structured and potentially parallelizable solution process. Saeed et al. \cite{saeed2016control} proposed two heuristic solutions (SRM and ITA) that decompose the resource allocation problem into control and data channel optimization.

    \item \textbf{Other Approaches:} The literature also showcases other techniques. Gang et al. \cite{gang2019inter} enhanced a greedy approach with randomness to explore the solution space further, while Cao et al. \cite{cao2021hybrid} utilized the K-shortest path algorithm for QKD network deployment.

\end{itemize}

\subsubsection{Hybrid Heuristics and Reinforcement Learning}
Hybrid Heuristics and Reinforcement Learning combine different heuristic approaches or use Reinforcement Learning(RL) to learn effective strategies. Hybrid Heuristics, such as Matheuristics, which combine metaheuristics with exact methods, and Hyperheuristics, which use a higher-level strategy to manage lower-level heuristics, leverage the strengths of different approaches. RL-based approaches use RL agents to learn decision-making policies. These methods can leverage the strengths of different approaches and RL can learn adaptive strategies, but they can be complex to design and RL training can be time-consuming.

In the context of 5G radio resource management, Boutiba et al. \cite{boutiba2023optimal} employed a Deep Reinforcement Learning based approach to tackle the NP-hard resource allocation problem in a mixed-numerology environment. Their reinforcement learning scheduler dynamically selects numerology and allocates Physical Resource Blocks (PRBs) to users, ensuring efficient resource utilization while meeting diverse QoS requirements. The authors demonstrate the scalability and real-time capabilities of their solution.

Similarly, Sharara et al. \cite{sharara2022policy} used policy gradient-based reinforcement learning to address the challenge of computing resource allocation in an O-RAN architecture. They modelled the allocation problem as an ILP problem and then employed an RL algorithm to learn optimal resource allocation policies. Their approach effectively balances the allocation of resources between eMBB and URLLC users, ensuring fairness and QoS optimization.

In NFV, Pham et al. (\cite{pham2021traffic} proposed a reinforcement learning-based algorithm to solve the traffic engineering problem for service function chaining with delay guarantees. The author formulated the problem as a MILP model and then employed reinforcement learning to find near-optimal solutions in large-scale and dynamic network environments.

\subsubsection{Approximation Algorithms}
Approximation Algorithms provide solutions guaranteed to be within a certain factor of the optimal solution. Methods include Greedy Approximation, which makes locally optimal choices with provable performance guarantees, LP Relaxation and Rounding, which relaxes integer constraints and rounds the solution, Primal-Dual Approximation, which uses duality theory, and Cutting Plane Methods, which add valid inequalities. These algorithms provide performance guarantees and can be computationally efficient, but the approximation factor might not be tight enough, and not all problems admit good approximation algorithms.

Among the selected studies, the works by Feng et al. \cite{feng2017joint} and Mharsi et al. \cite{mharsi2018joint} addressed resource allocation problems in wireless communication networks. The former focuses on maximizing the downlink rate in massive MIMO HetNets, while the latter aims to optimize communication latency and resource consumption in C-RANs. Both studies employ a combination of exact ILP models and approximation algorithms to achieve near-optimal solutions efficiently.

The papers by Almasaeid \cite{almasaeid2023minimum} and Salameh et al. \cite{salameh2022opportunistic} delved into spectrum allocation problems. The former investigated spectrum allocation in dynamic spectrum access networks with the goal of minimizing cost while meeting QoS requirements. The latter proposed a resource allocation protocol for future cellular networks that leverages CR technology and D-OFDMA to maximize network throughput. Both studies formulated the problems as ILPs and developed sub-optimal algorithms based on relaxation and rounding techniques.

The works by Sharara et al. \cite{sharara2023minimizing} and Bi et al. \cite{bi2019resource} focused on resource allocation in 5G networks. The former investigated energy consumption in C-RANs and proposed a joint allocation approach that simultaneously optimized radio and computing resources. The latter addressed the challenge of supporting ultra-low latency services in hierarchical 5G networks. Both studies utilized MILP models and developed approximation algorithms to achieve near-optimal solutions with reduced computational complexity.

Finally, the papers by Yang et al. \cite{yang2023space}, and Quang et al. \cite{quang2018qaav} explored VNF placement and resource allocation problems. The former investigated joint VNF deployment and flow routing in SINs to maximize the number of successfully completed missions. The latter addressed the dynamic adaptation and optimization of VNF placements in wireless networks with QoS constraints. Both studies formulated the problems as ILPs or MILPs and proposed novel approximation algorithms to achieve near-optimal solutions efficiently.

\subsubsection{Decomposition-Based Heuristics}
Decomposition-based heuristics decompose the problem into smaller subproblems, solve them independently, and combine the solutions. Benders' Decomposition decomposes the problem into a master problem and subproblems, iteratively solving them and exchanging information, while Dantzig-Wolfe's Decomposition uses column generation to solve the master problem. These methods can handle large-scale problems by exploiting problem structure and can be parallelized, but the solution quality depends on the decomposition strategy, and they may require specialized algorithms for the subproblems.

Marijanovic et al. \cite{marijanovic2020multiplexing} utilized the Dantzig-Wolfe decomposition, which partitioned the problem based on constraint structures, enabling parallel solving of sub-problems and coordination through a master problem. Yang et al. \cite{yang2020towards} employed column generation methods, decomposing the problem into a master problem for selecting configurations and pricing problems for generating new configurations. Lastly, Gkatzios et al. \cite{gkatzios2019efficiency} proposed heuristics employing function grouping and greedy allocation strategies, further simplifying the problem and achieving near-optimal solutions with reduced computational time. These diverse approaches highlighted the versatility and effectiveness of decomposition-based heuristics in tackling the challenges posed by ILP problems in network optimization. Jia et al. \cite{jia2021toward} first tackled the problem using Benders decomposition, which provides an optimal solution to the MILP. To improve the efficiency for larger networks, they introduced an acceleration algorithm. This method uses approximations in both the master and subproblems, allowing for quicker, though sometimes less accurate, solutions. Shokrnezhad et al. \cite{shokrnezhad2018joint} tackled MILP by employing Lagrangian relaxation to decompose the problem into solvable sub-problems, then used the subgradient method to iteratively find a near-optimal solution.


\subsubsection{Heuristic-Based Relaxations}
Heuristic-Based Relaxations relax some constraints to obtain an easier problem, solve it, and use the solution to guide the search for a feasible solution. Lagrangian Relaxation relaxes constraints by adding them to the objective function with Lagrange multipliers, while Surrogate Relaxation combines multiple constraints into a single one. These methods can provide good bounds on the optimal solution and can be used within other methods, but the solution quality depends on the relaxation strategy, and finding good Lagrange multipliers or surrogate constraints can be challenging.

Among the selected studies, only Yao et al. \cite{9929619} used heuristic-based relaxations. They addressed the complex problem of user access control and resource allocation in RAN slicing. The authors employed a heuristic-based relaxation to simplify the problem by relaxing integer constraints, transforming it into a more manageable convex optimization problem. The solution to this relaxed problem was an upper bound for the original problem.



\subsubsection{Problem-Specific Heuristics}
Finally, Problem-Specific Heuristics are tailored to specific problem domains or structures. They exploit unique characteristics to find efficient solutions.

Ferdosian et al. \cite{ferdosian20225g} tackled the challenge of allocating 5G NR resources to diverse services like URLLC and eMBB. Recognizing the NP-hard nature of the problem, they reformulated it with relaxed constraints and proposed a greedy heuristic inspired by bin packing optimization. This approach prioritized URLLC services while maximizing eMBB throughput, demonstrating the effectiveness of heuristics in balancing conflicting objectives.

Alsheyab et al. \cite{alsheyab2019near} investigated the association problem between NFPs and small cells in 5G+ networks. The formulated ILP aimed to maximize the total sum rate under QoS, bandwidth, and link constraints. The authors proposed two sub-optimal algorithms (centralized and distributed) based on weighted bipartite matching and stable marriage matching, respectively. These heuristics offered a trade-off between optimality and computational complexity, crucial in the dynamic environment of NFP-assisted networks.

Chiaraviglio et al. \cite{chiaraviglio20215g} addressed the 5G network planning problem considering both service coverage and electromagnetic field (EMF) constraints. The MILP formulation aimed to minimize installation costs and maximize service coverage while adhering to strict EMF regulations. The authors proposed the PLATEA heuristic, which employed a divide-and-conquer approach and leveraged linear constraints to reduce complexity. This demonstrated the adaptability of problem-specific heuristics to incorporate real-world regulatory constraints.

Through adaptive codebook design and assignment, Zhai et al. \cite{zhai2017adaptive} explored energy-saving in SCMA networks. The formulated ILP minimized detection complexity in the uplink and total power consumption in the downlink. The proposed DCSA and FIA algorithms, based on dual coordinate search and Lagrangian dual decomposition, showcased the potential of heuristics in cross-layer optimization, bridging the gap between physical and MAC layer design.

Zoha et al. \cite{zoha2017leveraging} proposed a proactive approach to energy saving and ICI mitigation in ultra-dense 5G networks. They identified predictable traffic patterns by analyzing real-world CDR data and formulated a joint optimization problem. The ECA heuristic leveraged this intelligence to schedule radio resources and small cell sleep cycles, achieving significant energy savings without compromising QoS. This highlighted the power of problem-specific heuristics in exploiting real-world data insights for network optimization.


\section{Discussions}
\label{sec:Discussion}
The survey paper presents a comprehensive overview of the application of LP, ILP, and MILP in optimizing resource allocation for 5G and B5G networks. The paper categorizes the literature based on network architectures, objective functions, resource types, constraints, and solution techniques, providing a structured analysis of the current research landscape. The following discussion synthesizes and analyzes the key findings and implications of this survey:




Due to their adaptability and effectiveness, LP, ILP, and MILP are widely utilized across various network architectures and resource allocation problems. The research reviewed demonstrates their broad applicability, highlighting their versatility in addressing complex challenges in modern network resource management. These optimization techniques handle a range of objectives, including network performance optimization, resource utilization efficiency, economic considerations, and reliability, while managing constraints such as capacity, QoS, energy efficiency, and interference.


The NP-hard nature of ILP and MILP necessitates a range of solution techniques, from exact methods and heuristics to approximation algorithms and hybrid approaches, each with its trade-offs between optimality and computational efficiency. The survey reveals a clear trend towards the use of more sophisticated and specialized solution techniques. While traditional solvers remain relevant for certain problem formulations, the increasing complexity of 5G and B5G networks has driven the adoption of advanced methods such as decomposition techniques, metaheuristics, approximation algorithms, and hybrid approaches that combine heuristics with machine learning. This trend reflects the need for computationally efficient and adaptable solutions that can handle large-scale and dynamic network environments. Integrating Artificial Intelligence (AI) and machine learning with optimization models is a promising trend, enabling dynamic and adaptive resource allocation strategies that respond to next-generation networks' complexities.

The following are the possible future directions of resource allocation problems using LP, ILP, and MILP models. 

\begin{itemize}

    \item The survey indicates a significant trend towards using AI  for solving optimization problems modeled as ILP and MILP. This integration promises to enhance the solution of complex optimization models in network resource allocation \cite{wu2024ai}. AI-driven approaches can offer more intelligent and adaptive strategies that respond dynamically to changing network conditions, and user demands can be used to enhance ILP and MILP models in various ways, making them more tractable for complex network optimization problems. AI-guided heuristics and metaheuristics can quickly be developed to find high-quality solutions, especially for large-scale ILP and MILP problems \cite{mirhosseini2023fast}. Machine learning techniques can be used for parameter tuning and algorithm selection in solvers, improving efficiency. Hybrid algorithms that combine mathematical programming with AI can further enhance the ability to solve complex problems more effectively. Lastly, AI can be utilized for real-time decision-making and adaptive resource allocation based on the outcomes of solved ILP and MILP models.

    \item Handling the complexities of 5G and B5G networks, especially in dense and ultra-dense deployments, requires distributed and parallel optimization approaches. Distributed algorithms leverage the decentralized nature of these networks, while parallel computing techniques, using edge and cloud resources, help solve large-scale ILP and MILP problems more efficiently. Decomposition, hierarchical approaches, and heuristics further break down these complex problems into manageable sub-problems, enabling near-optimal and scalable solutions in dynamic, large-scale environments \cite{khadidos2024distribution, bedda2023new, jaumard2023nested, arshad2023energy}.

    \item  A promising future direction is using quantum computing to optimize network operations in 5G and B5G environments. Quantum algorithms have the potential to solve large-scale ILP and MILP problems more efficiently than classical approaches \cite{fayad2024optimization, zhang2024quantum}. The exploration of hybrid quantum-classical algorithms also offers exciting possibilities, combining the computational strengths of both paradigms. This hybrid approach could significantly improve the efficiency and effectiveness of solving complex network optimization challenges, paving the way for enhanced performance in next-generation networks.

    \item A possible future direction is the integration of explainable AI techniques into AI-assisted ILP and MILPsolutions. This aims to enhance the transparency and interpretability of AI-driven optimization processes, making them more accessible to network operators and decision-makers. By clarifying how AI models arrive at their conclusions, explainability fosters trust in AI-generated solutions. Furthermore, developing advanced visualization tools will enable operators to better grasp complex, large-scale optimization outcomes, particularly in high-stakes environments. This will ensure that AI-driven decisions are actionable, reliable, and aligned with operational goals.

    \item A promising direction is the use of Large Language Models (LLMs) to automate or assist in generating optimization models from natural language descriptions. With increasing research in this area, methods have already been proposed to generate simple optimization problems from natural language inputs \cite{ahmed2024lm4opt, ahmaditeshnizi2023optimus, ramamonjison2023nl4opt}. This approach could simplify network optimization, making it more accessible to non-experts and speeding up the prototyping process. However, since network optimization requires domain-specific knowledge, LLMs must be capable of understanding and handling the relevant constraints. By training LLMs to interpret network resource allocation problems, they could extract key entities, objectives, and constraints, automatically generating mathematical models for LP, ILP, or MILP solvers. LLMs could also assist in validating these models, though expert input may still be necessary in complex scenarios to ensure the accuracy and sophistication of the generated models.

\end{itemize}

\section{Conclusions}
\label{sec:Conclusions}
This paper presents a comprehensive survey of 103 studies on resource allocation strategies in 5G and B5G networks, focusing on optimization problems modelled as LP, ILP, and MILP. The survey categorizes these studies based on network architectures, resource allocation types, objective functions, constraints, and solution techniques. The findings highlight the widespread adoption of LP, ILP, and MILP models across various network architectures and resource types, demonstrating their effectiveness in optimizing network performance, resource utilization, and energy efficiency. The survey also discusses solution methods for NP-hard ILP and MILP problems, categorizing them into different approaches. Additionally, it explores emerging trends, such as integrating AI and machine learning with optimization models, suggesting promising future research directions in network optimization. The paper concludes that LP, ILP, and MILP models have been extensively employed across various network architectures, resource types, objective functions, and constraints, and they remain critical to optimizing next-generation networks.


\begin{thebibliography}{100}

\bibitem{9355403}
Y.~Xu, G.~Gui, H.~Gacanin, and F.~Adachi, ``A survey on resource allocation for 5g heterogeneous networks: Current research, future trends, and challenges,'' {\em IEEE Communications Surveys \& Tutorials}, vol.~23, no.~2, pp.~668--695, 2021.

\bibitem{mahmood2021industrial}
A.~Mahmood, L.~Beltramelli, S.~F. Abedin, S.~Zeb, N.~I. Mowla, S.~A. Hassan, E.~Sisinni, and M.~Gidlund, ``Industrial iot in 5g-and-beyond networks: Vision, architecture, and design trends,'' {\em IEEE Transactions on Industrial Informatics}, vol.~18, no.~6, pp.~4122--4137, 2021.

\bibitem{mishra2020survey}
D.~Mishra and E.~Natalizio, ``A survey on cellular-connected uavs: Design challenges, enabling 5g/b5g innovations, and experimental advancements,'' {\em Computer Networks}, vol.~182, p.~107451, 2020.

\bibitem{ebrahimi2024resource}
S.~Ebrahimi, F.~Bouali, and O.~C. Haas, ``Resource management from single-domain 5g to end-to-end 6g network slicing: A survey,'' {\em IEEE Communications Surveys \& Tutorials}, 2024.

\bibitem{fayad2024optimization}
A.~Fayad, ``Optimization of 5g and beyond networks for cost-and energy-efficiency,'' 2024.

\bibitem{machiwa2024comprehensive}
E.~J. Machiwa, V.~G. Masanja, M.~F. Kisangiri, and J.~W. Matiko, ``A comprehensive survey on linear programming and energy optimization methods for maximizing lifetime of wireless sensor network,'' {\em Discover Computing}, vol.~27, no.~1, p.~21, 2024.

\bibitem{xu2021survey}
Y.~Xu, G.~Gui, H.~Gacanin, and F.~Adachi, ``A survey on resource allocation for 5g heterogeneous networks: Current research, future trends, and challenges,'' {\em IEEE Communications Surveys \& Tutorials}, vol.~23, no.~2, pp.~668--695, 2021.

\bibitem{ganian2017going}
R.~Ganian, S.~Ordyniak, and M.~Ramanujan, ``Going beyond primal treewidth for (m) ilp,'' in {\em Proceedings of the AAAI Conference on Artificial Intelligence}, vol.~31, 2017.

\bibitem{arnhold2024network}
F.~Arnhold, S.~S. Anbazhagan, L.~R. Prade, J.~M. Nogueira, A.~Klautau, and C.~B. Both, ``Network slicing support by fronthaul interface in disaggregated radio access networks: A survey,'' {\em IEEE Transactions on Network and Service Management}, 2024.

\bibitem{elijah2022intelligent}
O.~Elijah, S.~K.~A. Rahim, W.~K. New, C.~Y. Leow, K.~Cumanan, and T.~K. Geok, ``Intelligent massive mimo systems for beyond 5g networks: An overview and future trends,'' {\em IEEE Access}, vol.~10, pp.~102532--102563, 2022.

\bibitem{li20185g}
Z.~Li, M.~A. Uusitalo, H.~Shariatmadari, and B.~Singh, ``5g urllc: Design challenges and system concepts,'' in {\em 2018 15th international symposium on wireless communication systems (ISWCS)}, pp.~1--6, IEEE, 2018.

\bibitem{mirhosseini2023fast}
M.~Mirhosseini, M.~Fazlali, M.~K. Fallah, and J.-A. Lee, ``A fast milp solver for high-level synthesis based on heuristic model reduction and enhanced branch and bound algorithm,'' {\em The Journal of Supercomputing}, vol.~79, no.~11, pp.~12042--12073, 2023.

\bibitem{smith2012measuring}
K.~Smith-Miles and L.~Lopes, ``Measuring instance difficulty for combinatorial optimization problems,'' {\em Computers \& Operations Research}, vol.~39, no.~5, pp.~875--889, 2012.

\bibitem{bixby1992very}
R.~E. Bixby, J.~W. Gregory, I.~J. Lustig, R.~E. Marsten, and D.~F. Shanno, ``Very large-scale linear programming: A case study in combining interior point and simplex methods,'' {\em Operations Research}, vol.~40, no.~5, pp.~885--897, 1992.

\bibitem{wiese2016interplay}
S.~Wiese, ``On the interplay of mixed integer linear, mixed integer nonlinear and constraint programming,'' 2016.

\bibitem{baghestani2023logistics}
A.~Baghestani, M.~Abbasi, S.~Rastegar, A.~R. Mamdoohi, A.~Afaghpoor, and M.~Saffarzadeh, ``Logistics village location with capacity planning problem, an milp model approach,'' {\em Sustainability}, vol.~15, no.~5, p.~4633, 2023.

\bibitem{amaldi2011milp}
E.~Amaldi, A.~Capone, L.~G. Gianoli, and L.~Mascetti, ``A milp-based heuristic for energy-aware traffic engineering with shortest path routing,'' in {\em International Conference on Network Optimization}, pp.~464--477, Springer, 2011.

\bibitem{meng2020milp}
L.~Meng, Y.~Ren, B.~Zhang, J.-Q. Li, H.~Sang, and C.~Zhang, ``Milp modeling and optimization of energy-efficient distributed flexible job shop scheduling problem,'' {\em IEEE Access}, vol.~8, pp.~191191--191203, 2020.

\bibitem{marsch20165g}
P.~Marsch, I.~Da~Silva, O.~Bulakci, M.~Tesanovic, S.~E. El~Ayoubi, T.~Rosowski, A.~Kaloxylos, and M.~Boldi, ``5g radio access network architecture: Design guidelines and key considerations,'' {\em IEEE Communications Magazine}, vol.~54, no.~11, pp.~24--32, 2016.

\bibitem{habibi2019comprehensive}
M.~A. Habibi, M.~Nasimi, B.~Han, and H.~D. Schotten, ``A comprehensive survey of ran architectures toward 5g mobile communication system,'' {\em Ieee Access}, vol.~7, pp.~70371--70421, 2019.

\bibitem{rodoshi2020resource}
R.~T. Rodoshi, T.~Kim, and W.~Choi, ``Resource management in cloud radio access network: Conventional and new approaches,'' {\em Sensors}, vol.~20, no.~9, p.~2708, 2020.

\bibitem{hasabelnaby2018joint}
M.~A. Hasabelnaby, H.~A. Selmy, and M.~I. Dessouky, ``Joint optimal transceiver placement and resource allocation schemes for redirected cooperative hybrid fso/mmw 5g fronthaul networks,'' {\em Journal of Optical Communications and Networking}, vol.~10, no.~12, pp.~975--990, 2018.

\bibitem{sharara2023minimizing}
M.~Sharara, F.~Fossati, S.~Hoteit, V.~V{\`e}que, and F.~Bassi, ``Minimizing energy consumption by joint radio and computing resource allocation in cloud-ran,'' {\em Computer Networks}, vol.~234, p.~109870, 2023.

\bibitem{gkatzios2018compute}
N.~Gkatzios, M.~Anastasopoulos, A.~Tzanakaki, and D.~Simeonidou, ``Compute resource disaggregation: An enabler for efficient 5g ran softwarisation,'' in {\em 2018 European Conference on Networks and Communications (EuCNC)}, pp.~1--5, IEEE, 2018.

\bibitem{gkatzios2019efficiency}
N.~Gkatzios, M.~Anastasopoulos, A.~Tzanakaki, and D.~Simeonidou, ``Efficiency gains in 5g softwarised radio access networks,'' {\em EURASIP Journal on Wireless Communications and Networking}, vol.~2019, pp.~1--11, 2019.

\bibitem{de2019optimal}
A.~De~Domenico, Y.-F. Liu, and W.~Yu, ``Optimal computational resource allocation and network slicing deployment in 5g hybrid c-ran,'' in {\em ICC 2019-2019 IEEE International Conference on Communications (ICC)}, pp.~1--6, IEEE, 2019.

\bibitem{ali2017joint}
M.~Ali, Q.~Rabbani, M.~Naeem, S.~Qaisar, and F.~Qamar, ``Joint user association, power allocation, and throughput maximization in 5g h-cran networks,'' {\em IEEE Transactions on Vehicular Technology}, vol.~66, no.~10, pp.~9254--9262, 2017.

\bibitem{fayad2022design}
A.~Fayad, T.~Cinkler, J.~Rak, and M.~Jha, ``Design of cost-efficient optical fronthaul for 5g/6g networks: An optimization perspective,'' {\em Sensors}, vol.~22, no.~23, p.~9394, 2022.

\bibitem{kazemifard2021minimum}
N.~Kazemifard and V.~Shah-Mansouri, ``Minimum delay function placement and resource allocation for open ran (o-ran) 5g networks,'' {\em Computer Networks}, vol.~188, p.~107809, 2021.

\bibitem{9827120}
Z.~Sasan and S.~Khorsandi, ``Slice-aware resource calendaring in cloud-based radio access networks,'' in {\em 2022 30th International Conference on Electrical Engineering (ICEE)}, pp.~1005--1009, 2022.

\bibitem{tang2018queue}
L.~Tang, G.~Zhao, C.~Wang, P.~Zhao, and Q.~Chen, ``Queue-aware reliable embedding algorithm for 5g network slicing,'' {\em Computer networks}, vol.~146, pp.~138--150, 2018.

\bibitem{mharsi2018joint}
N.~Mharsi and M.~Hadji, ``Joint optimization of communication latency and resource allocation in cloud radio access networks,'' in {\em 2018 International Conference on Smart Communications in Network Technologies (SaCoNeT)}, pp.~13--18, IEEE, 2018.

\bibitem{arouk2017multi}
O.~Arouk, N.~Nikaein, and T.~Turletti, ``Multi-objective placement of virtual network function chains in 5g,'' in {\em 2017 IEEE 6th International Conference on Cloud Networking (CloudNet)}, pp.~1--6, IEEE, 2017.

\bibitem{klinkowski2022dedicated}
M.~Klinkowski and M.~Jaworski, ``Dedicated path protection with wavelength aggregation in 5g packet-optical xhaul access networks,'' {\em Journal of Lightwave Technology}, vol.~41, no.~6, pp.~1591--1602, 2022.

\bibitem{basu2020reverse}
D.~Basu, A.~Jain, U.~Ghosh, and R.~Datta, ``A reverse path-flow mechanism for latency aware controller placement in vsdn enabled 5g network,'' {\em IEEE Transactions on Industrial Informatics}, vol.~17, no.~10, pp.~6885--6893, 2020.

\bibitem{mounika2021downlink}
R.~Mounika, A.~Kumar, K.~Kuchi, {\em et~al.}, ``Downlink resource allocation for 5g-nr massive mimo systems,'' in {\em 2021 National Conference on Communications (NCC)}, pp.~1--6, IEEE, 2021.

\bibitem{xue2023joint}
T.~Xue, K.~A. Memon, and C.~Li, ``Joint resource allocation in twdm-pon-enabled cell-free mmimo system,'' in {\em Photonics}, vol.~10, p.~1180, MDPI, 2023.

\bibitem{feng2017adaptive}
M.~Feng and S.~Mao, ``Adaptive pilot design for massive mimo hetnets with wireless backhaul,'' in {\em 2017 14th Annual IEEE International Conference on Sensing, Communication, and Networking (SECON)}, pp.~1--9, IEEE, 2017.

\bibitem{feng2017joint}
M.~Feng, S.~Mao, and T.~Jiang, ``Joint frame design, resource allocation and user association for massive mimo heterogeneous networks with wireless backhaul,'' {\em IEEE Transactions on Wireless Communications}, vol.~17, no.~3, pp.~1937--1950, 2017.

\bibitem{zoha2017leveraging}
A.~Zoha, A.~Saeed, H.~Farooq, A.~Rizwan, A.~Imran, and M.~A. Imran, ``Leveraging intelligence from network cdr data for interference aware energy consumption minimization,'' {\em IEEE Transactions on Mobile Computing}, vol.~17, no.~7, pp.~1569--1582, 2017.

\bibitem{you2017load}
L.~You and D.~Yuan, ``Load optimization with user association in cooperative and load-coupled lte networks,'' {\em IEEE Transactions on Wireless Communications}, vol.~16, no.~5, pp.~3218--3231, 2017.

\bibitem{hadi2020patient}
M.~S. Hadi, A.~Q. Lawey, T.~E. El-Gorashi, and J.~M. Elmirghani, ``Patient-centric hetnets powered by machine learning and big data analytics for 6g networks,'' {\em IEEE Access}, vol.~8, pp.~85639--85655, 2020.

\bibitem{swetha2018dynamic}
G.~D. Swetha, J.~Grover, and G.~R. Murthy, ``Dynamic channel allocation in small cells,'' in {\em 2018 7th International Conference on Reliability, Infocom Technologies and Optimization (Trends and Future Directions)(ICRITO)}, pp.~582--588, IEEE, 2018.

\bibitem{xie2017energy}
R.~Xie, Z.~Li, T.~Huang, and Y.~Liu, ``Energy-efficient joint content caching and small base station activation mechanism design in heterogeneous cellular networks,'' {\em China Communications}, vol.~14, no.~10, pp.~70--83, 2017.

\bibitem{saghezchi2017energy}
F.~B. Saghezchi, A.~Radwan, and J.~Rodriguez, ``Energy-aware relay selection in cooperative wireless networks: An assignment game approach,'' {\em Ad Hoc Networks}, vol.~56, pp.~96--108, 2017.

\bibitem{liu2020joint}
J.-S. Liu, C.-H.~R. Lin, and Y.-C. Hu, ``Joint resource allocation, user association, and power control for 5g lte-based heterogeneous networks,'' {\em IEEE Access}, vol.~8, pp.~122654--122672, 2020.

\bibitem{jain2021user}
A.~Jain, E.~Lopez-Aguilera, and I.~Demirkol, ``User association and resource allocation in 5g (aura-5g): A joint optimization framework,'' {\em Computer Networks}, vol.~192, p.~108063, 2021.

\bibitem{jalali2020joint}
A.~Jalali and Z.~Ding, ``Joint detection and decoding of polar coded 5g control channels,'' {\em IEEE Transactions on Wireless Communications}, vol.~19, no.~3, pp.~2066--2078, 2020.

\bibitem{marijanovic2020multiplexing}
L.~Marijanovi{\'c}, S.~Schwarz, and M.~Rupp, ``Multiplexing services in 5g and beyond: Optimal resource allocation based on mixed numerology and mini-slots,'' {\em IEEE Access}, vol.~8, pp.~209537--209555, 2020.

\bibitem{fayad20235g}
A.~Fayad, T.~Cinkler, and J.~Rak, ``5g millimeter wave network optimization: Dual connectivity and power allocation strategy,'' {\em IEEE Access}, 2023.

\bibitem{zhai2017adaptive}
D.~Zhai, ``Adaptive codebook design and assignment for energy saving in scma networks,'' {\em IEEE Access}, vol.~5, pp.~23550--23562, 2017.

\bibitem{salameh2022opportunistic}
H.~B. Salameh, H.~Al-Obiedollah, R.~Mahasees, and Y.~Jararweh, ``Opportunistic non-contiguous ofdma scheduling framework for future b5g/6g cellular networks,'' {\em Simulation Modelling Practice and Theory}, vol.~119, p.~102563, 2022.

\bibitem{sasikumar2021spectral}
S.~Sasikumar and J.~Jayakumari, ``Spectral-energy efficiencytradeoff enhancement: an optimal resource allocation framework for 5g underlay cognitive radio network,'' in {\em IEEE EUROCON 2021-19th International Conference on Smart Technologies}, pp.~284--289, IEEE, 2021.

\bibitem{zorello2022power}
L.~M.~M. Zorello, M.~Sodano, S.~Troia, and G.~Maier, ``Power-efficient baseband-function placement in latency-constrained 5g metro access,'' {\em IEEE Transactions on Green Communications and Networking}, vol.~6, no.~3, pp.~1683--1696, 2022.

\bibitem{sharara2022policy}
M.~Sharara, T.~Pamuklu, S.~Hoteit, V.~V{\`e}que, and M.~Erol-Kantarci, ``Policy-gradient-based reinforcement learning for computing resources allocation in o-ran,'' in {\em 2022 IEEE 11th International Conference on Cloud Networking (CloudNet)}, pp.~229--236, IEEE, 2022.

\bibitem{javad2021re}
M.~Javad-Kalbasi and S.~Valaee, ``Re-configuration of uav relays in 6g networks,'' in {\em 2021 IEEE International Conference on Communications Workshops (ICC Workshops)}, pp.~1--6, IEEE, 2021.

\bibitem{alwan2018joint}
S.~Alwan, I.~Fajjari, and N.~Aitsaadi, ``Joint multicast routing and ofdm resource allocation in lte-d2d 5g cellular network,'' in {\em NOMS 2018-2018 IEEE/IFIP Network Operations and Management Symposium}, pp.~1--9, IEEE, 2018.

\bibitem{yoon2018resource}
T.~Yoon, T.~H. Nguyen, X.~T. Nguyen, D.~Yoo, B.~Jang, {\em et~al.}, ``Resource allocation for noma-based d2d systems coexisting with cellular networks,'' {\em IEEE Access}, vol.~6, pp.~66293--66304, 2018.

\bibitem{vlachos2017moca}
C.~Vlachos and V.~Friderikos, ``Moca: Multiobjective cell association for device-to-device communications,'' {\em IEEE Transactions on Vehicular Technology}, vol.~66, no.~10, pp.~9313--9327, 2017.

\bibitem{8676260}
D.~Sattar and A.~Matrawy, ``Optimal slice allocation in 5g core networks,'' {\em IEEE Networking Letters}, vol.~1, no.~2, pp.~48--51, 2019.

\bibitem{alsheyab2019near}
H.~Y. Alsheyab, S.~Choudhury, E.~Bedeer, and S.~S. Ikki, ``Near-optimal resource allocation algorithms for 5g+ cellular networks,'' {\em IEEE Transactions on Vehicular Technology}, vol.~68, no.~7, pp.~6578--6592, 2019.

\bibitem{chiaraviglio20215g}
L.~Chiaraviglio, C.~Di~Paolo, and N.~Blefari-Melazzi, ``5g network planning under service and emf constraints: Formulation and solutions,'' {\em IEEE Transactions on Mobile Computing}, vol.~21, no.~9, pp.~3053--3070, 2021.

\bibitem{zeng2019energy}
Y.~Zeng, A.~Al-Quzweeni, T.~E. El-Gorashi, and J.~M. Elmirghani, ``Energy efficient virtualization framework for 5g f-ran,'' in {\em 2019 21st International Conference on Transparent Optical Networks (ICTON)}, pp.~1--4, IEEE, 2019.

\bibitem{al2019optimized}
A.~N. Al-Quzweeni, A.~Q. Lawey, T.~E. Elgorashi, and J.~M. Elmirghani, ``Optimized energy aware 5g network function virtualization,'' {\em Ieee Access}, vol.~7, pp.~44939--44958, 2019.

\bibitem{spinelli2022migration}
F.~Spinelli and V.~Mancuso, ``A migration path toward green edge gaming,'' in {\em 2022 IEEE 23rd International Symposium on a World of Wireless, Mobile and Multimedia Networks (WoWMoM)}, pp.~347--356, IEEE, 2022.

\bibitem{chu2023hirn}
Z.~Chu, L.~Guo, J.~Wang, Q.~Qi, Z.~Zhuang, H.~Sun, and C.~Zhou, ``Hirn: A hierarchical intent refinement approach for dependable network slicing with multi-path resource allocation,'' in {\em 2023 53rd Annual IEEE/IFIP International Conference on Dependable Systems and Networks-Supplemental Volume (DSN-S)}, pp.~7--13, IEEE, 2023.

\bibitem{deb2022fog}
P.~Deb and D.~De, ``Fog computing-based beam allocation and co-operative task distribution model for green 5g mobile network,'' {\em Innovations in Systems and Software Engineering}, pp.~1--12, 2022.

\bibitem{gatzianas2021energy}
M.~Gatzianas, A.~Mesodiakaki, G.~Kalfas, and N.~Pleros, ``Energy-efficient joint computational and network resource planning in beyond 5g networks,'' in {\em 2021 IEEE Global Communications Conference (GLOBECOM)}, pp.~1--6, IEEE, 2021.

\bibitem{lagkas2021optimized}
T.~Lagkas, D.~Klonidis, P.~Sarigiannidis, and I.~Tomkos, ``Optimized joint allocation of radio, optical, and mec resources for the 5g and beyond fronthaul,'' {\em IEEE Transactions on Network and Service Management}, vol.~18, no.~4, pp.~4639--4653, 2021.

\bibitem{pan2020multi}
Y.~Pan, L.~Gao, J.~Luo, T.~Wang, and J.~Luo, ``A multi-dimensional resource crowdsourcing framework for mobile edge computing,'' in {\em ICC 2020-2020 IEEE International Conference on Communications (ICC)}, pp.~1--7, IEEE, 2020.

\bibitem{ge2021dynamic}
L.~Ge, J.~Zhou, and Z.~Zheng, ``Dynamic hierarchical caching resource allocation for 5g-icn slice,'' {\em IEEE Access}, vol.~9, pp.~134972--134983, 2021.

\bibitem{kim2022modems}
T.~Kim, S.~D. Sathyanarayana, S.~Chen, Y.~Im, X.~Zhang, S.~Ha, and C.~Joe-Wong, ``Modems: Optimizing edge computing migrations for user mobility,'' {\em IEEE Journal on Selected Areas in Communications}, vol.~41, no.~3, pp.~675--689, 2022.

\bibitem{jia2021toward}
Z.~Jia, M.~Sheng, J.~Li, and Z.~Han, ``Toward data collection and transmission in 6g space--air--ground integrated networks: Cooperative hap and leo satellite schemes,'' {\em IEEE Internet of Things Journal}, vol.~9, no.~13, pp.~10516--10528, 2021.

\bibitem{yang2020towards}
Y.~Yang, X.~Chang, Z.~Jia, Z.~Han, and Z.~Han, ``Towards 6g joint haps-mec-cloud 3c resource allocation for delay-aware computation offloading,'' in {\em 2020 IEEE Intl Conf on Parallel \& Distributed Processing with Applications, Big Data \& Cloud Computing, Sustainable Computing \& Communications, Social Computing \& Networking (ISPA/BDCloud/SocialCom/SustainCom)}, pp.~175--182, IEEE, 2020.

\bibitem{yang2023space}
H.~Yang, W.~Liu, J.~Li, and T.~Q. Quek, ``Space information network with joint virtual network function deployment and flow routing strategy with qos constraints,'' {\em IEEE Journal on Selected Areas in Communications}, 2023.

\bibitem{monteiro2018distributed}
V.~F. Monteiro, D.~A. Sousa, T.~F. Maciel, F.~R.~P. Cavalcanti, C.~F. e~Silva, and E.~B. Rodrigues, ``Distributed rrm for 5g multi-rat multiconnectivity networks,'' {\em IEEE Systems Journal}, vol.~13, no.~1, pp.~192--203, 2018.

\bibitem{saeed2016control}
A.~Saeed, E.~Katranaras, M.~Dianati, and M.~A. Imran, ``Control and data channel resource allocation in ofdma heterogeneous networks,'' {\em Journal of Signal Processing Systems}, vol.~85, pp.~183--199, 2016.

\bibitem{guo2023delay}
Z.~Guo, X.~Ji, W.~You, M.~Xu, Y.~Zhao, Z.~Cheng, and D.~Zhou, ``Delay optimal for reliability-guaranteed concurrent transmissions with raptor code in multi-access 6g edge network,'' {\em Computer Networks}, vol.~228, p.~109716, 2023.

\bibitem{spinelli2022edge}
F.~Spinelli, A.~Bazco-Nogueras, and V.~Mancuso, ``Edge gaming: A greening perspective,'' {\em Computer Communications}, vol.~192, pp.~89--105, 2022.

\bibitem{gao2019deep}
M.~Gao, W.~Cui, D.~Gao, R.~Shen, J.~Li, and Y.~Zhou, ``Deep neural network task partitioning and offloading for mobile edge computing,'' in {\em 2019 IEEE Global Communications Conference (GLOBECOM)}, pp.~1--6, IEEE, 2019.

\bibitem{yue2023delay}
Y.~Yue, X.~Tang, W.~Yang, X.~Zhang, Z.~Zhang, C.~Gao, and L.~Xu, ``Delay-aware and resource-efficient vnf placement in 6g non-terrestrial networks,'' in {\em 2023 IEEE Wireless Communications and Networking Conference (WCNC)}, pp.~1--6, IEEE, 2023.

\bibitem{prastowo2022migration}
T.~Prastowo, A.~Shah, L.~Palopoli, R.~Passerone, and G.~Piro, ``Migration-aware optimized resource allocation in b5g edge networks,'' in {\em 2022 IEEE 19th Annual Consumer Communications \& Networking Conference (CCNC)}, pp.~106--113, IEEE, 2022.

\bibitem{ko2022pdras}
H.~Ko, J.~Lee, and S.~Pack, ``Pdras: Priority-based dynamic resource allocation scheme in 5g network slicing,'' {\em Journal of Network and Systems Management}, vol.~30, no.~4, p.~68, 2022.

\bibitem{ko2021priority}
H.~Ko, J.~Lee, and S.~Pack, ``Priority-based dynamic resource allocation scheme in network slicing,'' in {\em 2021 International Conference on Information Networking (ICOIN)}, pp.~62--64, IEEE, 2021.

\bibitem{li2019latency}
W.~Li, Y.~Zi, L.~Feng, F.~Zhou, P.~Yu, and X.~Qiu, ``Latency-optimal virtual network functions resource allocation for 5g backhaul transport network slicing,'' {\em Applied Sciences}, vol.~9, no.~4, p.~701, 2019.

\bibitem{boutiba2023optimal}
K.~Boutiba, M.~Bagaa, and A.~Ksentini, ``Optimal radio resource management in 5g nr featuring network slicing,'' {\em Computer Networks}, vol.~234, p.~109937, 2023.

\bibitem{mei20205g}
C.~Mei, J.~Liu, J.~Li, L.~Zhang, and M.~Shao, ``5g network slices embedding with sharable virtual network functions,'' {\em Journal of Communications and Networks}, vol.~22, no.~5, pp.~415--427, 2020.

\bibitem{harutyunyan2019orchestrating}
D.~Harutyunyan, R.~Fedrizzi, N.~Shahriar, R.~Boutaba, and R.~Riggio, ``Orchestrating end-to-end slices in 5g networks,'' in {\em 2019 15th International Conference on Network and Service Management (CNSM)}, pp.~1--9, IEEE, 2019.

\bibitem{yang2019two}
G.~Yang, Q.~Liu, X.~Zhou, Y.~Qian, and W.~Wu, ``Two-tier resource allocation in dynamic network slicing paradigm with deep reinforcement learning,'' in {\em 2019 IEEE Global Communications Conference (GLOBECOM)}, pp.~1--6, IEEE, 2019.

\bibitem{fendt2018formal}
A.~Fendt, C.~Mannweiler, L.~C. Schmelz, and B.~Bauer, ``A formal optimization model for 5g mobile network slice resource allocation,'' in {\em 2018 IEEE 9th Annual Information Technology, Electronics and Mobile Communication Conference (IEMCON)}, pp.~101--106, IEEE, 2018.

\bibitem{fendt2018network}
A.~Fendt, S.~Lohmuller, L.~C. Schmelz, and B.~Bauer, ``A network slice resource allocation and optimization model for end-to-end mobile networks,'' in {\em 2018 IEEE 5G World Forum (5GWF)}, pp.~262--267, IEEE, 2018.

\bibitem{jia2017efficient}
Q.~Jia, R.~Xie, T.~Huang, J.~Liu, and Y.~Liu, ``Efficient caching resource allocation for network slicing in 5g core network,'' {\em IET Communications}, vol.~11, no.~18, pp.~2792--2799, 2017.

\bibitem{papagianni2018rethinking}
C.~Papagianni, P.~Papadimitriou, and J.~S. Baras, ``Rethinking service chain embedding for cellular network slicing,'' in {\em 2018 IFIP Networking Conference (IFIP Networking) and Workshops}, pp.~1--9, IEEE, 2018.

\bibitem{de2020optimal}
A.~De~Domenico, Y.-F. Liu, and W.~Yu, ``Optimal virtual network function deployment for 5g network slicing in a hybrid cloud infrastructure,'' {\em IEEE Transactions on Wireless Communications}, vol.~19, no.~12, pp.~7942--7956, 2020.

\bibitem{morais2022placeran}
F.~Z. Morais, G.~M.~F. De~Almeida, L.~L. Pinto, K.~Cardoso, L.~M. Contreras, R.~da~Rosa~Righi, and C.~B. Both, ``Placeran: optimal placement of virtualized network functions in beyond 5g radio access networks,'' {\em IEEE Transactions on Mobile Computing}, 2022.

\bibitem{bi2019resource}
Y.~Bi, C.~Colman-Meixner, R.~Wang, F.~Meng, R.~Nejabati, and D.~Simeonidou, ``Resource allocation for ultra-low latency virtual network services in hierarchical 5g network,'' in {\em ICC 2019-2019 IEEE international conference on communications (ICC)}, pp.~1--7, IEEE, 2019.

\bibitem{yu2019network}
Y.~Yu, X.~Bu, K.~Yang, H.~K. Nguyen, and Z.~Han, ``Network function virtualization resource allocation based on joint benders decomposition and admm,'' {\em IEEE Transactions on Vehicular Technology}, vol.~69, no.~2, pp.~1706--1718, 2019.

\bibitem{biallach2022optimization}
H.~Biallach, M.~Bouhtou, and D.~Nace, ``Optimization of the virtual network function reconfiguration plan in 5g network slicing,'' in {\em ICN International Conference on Networks}, 2022.

\bibitem{mushtaq2023optimal}
M.~Mushtaq, M.~Golkarifard, N.~Shahriar, R.~Boutaba, and A.~Saleh, ``Optimal functional splitting, placement and routing for isolation-aware network slicing in ng-ran,'' in {\em 2023 19th International Conference on Network and Service Management (CNSM)}, pp.~1--5, IEEE, 2023.

\bibitem{gholami2023mobile}
A.~Gholami, N.~Torkzaban, and J.~S. Baras, ``Mobile network slicing under demand uncertainty: A stochastic programming approach,'' in {\em 2023 IEEE 9th International Conference on Network Softwarization (NetSoft)}, pp.~272--276, IEEE, 2023.

\bibitem{9379507}
J.~Gao, F.~Zhou, G.~Sun, L.~Feng, W.~Li, X.~Qiu, and X.~Chen, ``Resource allocation for network slices with delay-sensitive multimedia services,'' in {\em 2020 IEEE International Symposium on Broadband Multimedia Systems and Broadcasting (BMSB)}, pp.~1--5, 2020.

\bibitem{luu2021uncertainty}
Q.-T. Luu, S.~Kerboeuf, and M.~Kieffer, ``Uncertainty-aware resource provisioning for network slicing,'' {\em IEEE Transactions on Network and Service Management}, vol.~18, no.~1, pp.~79--93, 2021.

\bibitem{10056963}
H.~Li, Z.~Kong, Y.~Chen, L.~Wang, Z.~Lu, X.~Wen, W.~Jing, and W.~Xiang {\em IEEE Transactions on Network and Service Management}, vol.~20, no.~3, pp.~3652--3672, 2023.

\bibitem{9929619}
J.~Yao, D.~Chen, Y.~Yu, and W.~Wang, ``Ran slice access control scheme based on qos and maximum network utility,'' in {\em 2022 IEEE 6th Advanced Information Technology, Electronic and Automation Control Conference (IAEAC )}, pp.~1853--1858, 2022.

\bibitem{9685847}
W.~Ben-Ameur, L.~Cano, and T.~Chahed, ``A framework for joint admission control, resource allocation and pricing for network slicing in 5g,'' in {\em 2021 IEEE Global Communications Conference (GLOBECOM)}, pp.~1--6, 2021.

\bibitem{ferdosian20225g}
N.~Ferdosian, S.~Berri, and A.~Chorti, ``5g new radio resource allocation optimization for heterogeneous services,'' in {\em 2022 International Symposium ELMAR}, pp.~1--6, IEEE, 2022.

\bibitem{you2018resource}
L.~You, Q.~Liao, N.~Pappas, and D.~Yuan, ``Resource optimization with flexible numerology and frame structure for heterogeneous services,'' {\em IEEE Communications Letters}, vol.~22, no.~12, pp.~2579--2582, 2018.

\bibitem{gatzianas2021offline}
M.~Gatzianas, A.~Mesodiakaki, G.~Kalfas, N.~Pleros, F.~Moscatelli, G.~Landi, N.~Ciulli, and L.~Lossi, ``Offline joint network and computational resource allocation for energy-efficient 5g and beyond networks,'' {\em Applied Sciences}, vol.~11, no.~22, p.~10547, 2021.

\bibitem{erbati2021application}
M.~M. Erbati and G.~Schiele, ``Application-and reliability-aware service function chaining to support low-latency applications in an nfv-enabled network,'' in {\em 2021 IEEE Conference on Network Function Virtualization and Software Defined Networks (NFV-SDN)}, pp.~120--123, IEEE, 2021.

\bibitem{gang2019inter}
J.~Gang and V.~Friderikos, ``Inter-tenant resource sharing and power allocation in 5g virtual networks,'' {\em IEEE Transactions on Vehicular Technology}, vol.~68, no.~8, pp.~7931--7943, 2019.

\bibitem{pham2021traffic}
T.-M. Pham, ``Traffic engineering based on reinforcement learning for service function chaining with delay guarantee,'' {\em IEEE Access}, vol.~9, pp.~121583--121592, 2021.

\bibitem{spinnewyn2018coordinated}
B.~Spinnewyn, P.~H. Isolani, C.~Donato, J.~F. Botero, and S.~Latr{\'e}, ``Coordinated service composition and embedding of 5g location-constrained network functions,'' {\em IEEE Transactions on Network and Service Management}, vol.~15, no.~4, pp.~1488--1502, 2018.

\bibitem{quang2018qaav}
P.~T.~A. Quang, K.~D. Singh, A.~Bradai, and A.~Benslimane, ``Qaav: Quality of service-aware adaptive allocation of virtual network functions in wireless network,'' in {\em 2018 IEEE International Conference on Communications (ICC)}, pp.~1--6, IEEE, 2018.

\bibitem{almasaeid2023minimum}
H.~M. Almasaeid, ``Minimum cost spectrum allocation with qos guarantees in multi-interface multi-hop dynamic spectrum access networks,'' {\em Computer Networks}, vol.~231, p.~109824, 2023.

\bibitem{8204056}
R.~P. Singh and G.~R. Murthy, ``Economic node allocation in software defined wireless networks with forecasted traffic and distance constraints,'' in {\em 2017 8th International Conference on Computing, Communication and Networking Technologies (ICCCNT)}, pp.~1--5, 2017.

\bibitem{al2021joint}
F.~N. Al-Wesabi, I.~Khan, M.~Alamgeer, A.~M. Al-Sharafi, B.~J. Choi, and A.~Aldosary, ``A joint algorithm for resource allocation in d2d 5g wireless networks,'' {\em Comput. Mater. Continua}, vol.~69, no.~1, pp.~301--317, 2021.

\bibitem{aletri2020effect}
O.~Z. Aletri, K.~D. Alazwary, S.~O. Saeed, S.~H. Mohamed, T.~E. El-Gorashi, M.~T. Alresheedi, and J.~M. Elmirghani, ``Effect of receiver orientation on resource allocation in optical wireless systems,'' in {\em 2020 22nd International Conference on Transparent Optical Networks (ICTON)}, pp.~1--6, IEEE, 2020.

\bibitem{aletri2020optimum}
O.~Z. Aletri, A.~A. Alahmadi, S.~O. Saeed, S.~H. Mohamed, T.~E. El-Gorashi, M.~T. Alresheedi, and J.~M. Elmirghani, ``Optimum resource allocation in 6g optical wireless communication systems,'' in {\em 2020 2nd 6G Wireless Summit (6G SUMMIT)}, pp.~1--6, IEEE, 2020.

\bibitem{cao2021hybrid}
Y.~Cao, Y.~Zhao, J.~Li, R.~Lin, J.~Zhang, and J.~Chen, ``Hybrid trusted/untrusted relay-based quantum key distribution over optical backbone networks,'' {\em IEEE Journal on Selected Areas in Communications}, vol.~39, no.~9, pp.~2701--2718, 2021.

\bibitem{kokkinos2019pattern}
P.~Kokkinos, P.~Soumplis, and E.~A. Varvarigos, ``Pattern-driven resource allocation in optical networks,'' {\em IEEE Transactions on Network and Service Management}, vol.~16, no.~2, pp.~489--504, 2019.

\bibitem{shokrnezhad2018joint}
M.~Shokrnezhad and S.~Khorsandi, ``Joint power control and channel assignment in uplink iot networks: A non-cooperative game and auction based approach,'' {\em Computer communications}, vol.~118, pp.~1--13, 2018.

\bibitem{rayani2018slicing}
M.~Rayani, D.~Naboulsi, R.~Glitho, and H.~Elbiaze, ``Slicing virtualized epc-based 5g core network for content delivery,'' in {\em 2018 IEEE Symposium on Computers and Communications (ISCC)}, pp.~00726--00729, IEEE, 2018.

\bibitem{silva2023dynamic}
R.~S. Silva, W.~Pires, S.~L. Correa, A.~Oliveira, and K.~V. Cardoso, ``Dynamic resources allocation in non-3gpp iot networks involving uavs,'' in {\em 2023 IEEE 97th Vehicular Technology Conference (VTC2023-Spring)}, pp.~1--5, IEEE, 2023.

\bibitem{9068264}
D.~H. Kim, S.~M.~A. Kazmi, A.~Ndikumana, A.~Manzoor, W.~Saad, and C.~S. Hong, ``Distributed radio slice allocation in wireless network virtualization: Matching theory meets auctions,'' {\em IEEE Access}, vol.~8, pp.~73494--73507, 2020.

\bibitem{wu2024ai}
S.~Wu, N.~Chen, A.~Xiao, P.~Zhang, C.~Jiang, and W.~Zhang, ``Ai-empowered virtual network embedding: a comprehensive survey,'' {\em IEEE Communications Surveys \& Tutorials}, 2024.

\bibitem{khadidos2024distribution}
A.~O. Khadidos, H.~Manoharan, S.~Selvarajan, A.~O. Khadidos, A.~M. Alshareef, and M.~Altwijri, ``Distribution of resources beyond 5g networks with heterogeneous parallel processing and graph optimization algorithms,'' {\em Cluster Computing}, pp.~1--19, 2024.

\bibitem{bedda2023new}
K.~Bedda, {\em New paradigms of distributed ai for improving 5G-based network systems performance}.
\newblock PhD thesis, 2023.

\bibitem{jaumard2023nested}
B.~Jaumard and Q.~H. Duong, ``A nested decomposition model for reliable nfv 5g network slicing,'' {\em IEEE Transactions on Network and Service Management}, 2023.

\bibitem{arshad2023energy}
R.~Arshad, M.~Farooq-i Azam, R.~Muzzammel, A.~Ghani, and C.~H. See, ``Energy efficiency and throughput optimization in 5g heterogeneous networks,'' {\em Electronics}, vol.~12, no.~9, p.~2031, 2023.

\bibitem{zhang2024quantum}
Y.~Zhang, Y.~Gong, L.~Fan, Y.~Wang, Z.~Han, and Y.~Guo, ``Quantum-assisted joint caching and power allocation for integrated satellite-terrestrial networks,'' {\em IEEE Transactions on Network Science and Engineering}, 2024.

\bibitem{ahmed2024lm4opt}
T.~Ahmed and S.~Choudhury, ``Lm4opt: Unveiling the potential of large language models in formulating mathematical optimization problems,'' {\em INFOR: Information Systems and Operational Research}, pp.~1--14, 2024.

\bibitem{ahmaditeshnizi2023optimus}
A.~AhmadiTeshnizi, W.~Gao, and M.~Udell, ``Optimus: Optimization modeling using mip solvers and large language models,'' {\em arXiv preprint arXiv:2310.06116}, 2023.

\bibitem{ramamonjison2023nl4opt}
R.~Ramamonjison, T.~Yu, R.~Li, H.~Li, G.~Carenini, B.~Ghaddar, S.~He, M.~Mostajabdaveh, A.~Banitalebi-Dehkordi, Z.~Zhou, {\em et~al.}, ``Nl4opt competition: Formulating optimization problems based on their natural language descriptions,'' in {\em NeurIPS 2022 Competition Track}, pp.~189--203, PMLR, 2023.

\end{thebibliography}
\end{document}